\documentclass[useAMS,usenatbib]{mn2e}
\usepackage{natbib}
\usepackage{amsmath}
\usepackage{graphicx}
\usepackage{multirow}
\usepackage{url}
\numberwithin{equation}{section}
\title[ISM chemistry \& cooling - II. Shielded gas]{Non-equilibrium chemistry and cooling in the diffuse interstellar medium - II. Shielded gas}
\author[A. J. Richings, J. Schaye and B. D. Oppenheimer]{A. J. Richings$^{1}$, J. Schaye$^{1}$ and B. D. Oppenheimer$^{2}$\\
$^{1}$Leiden Observatory, Leiden University, PO Box 9513, 2300 RA Leiden, the Netherlands\\
$^{2}$Center for Astrophysics and Space Astronomy, Department of Astrophysical and Planetary Sciences, University of Colorado, \\
389 UCB, Boulder, CO 80309, USA}

\begin{document}

\date{Accepted 2014 May 26. Received 2014 May 26; in original form 2014 March 24}

\pagerange{\pageref{firstpage}--\pageref{lastpage}} \pubyear{2014}

\maketitle

\label{firstpage}

\begin{abstract} 

We extend the non-equilibrium model for the chemical and thermal evolution of diffuse interstellar gas presented in \citet{richings14} to account for shielding from the UV radiation field. We attenuate the photochemical rates by dust and by gas, including absorption by H\textsc{i}, H$_{2}$, He\textsc{i}, He\textsc{ii} and CO where appropriate. We then use this model to investigate the dominant cooling and heating processes in interstellar gas as it becomes shielded from the UV radiation. We consider a one-dimensional plane-parallel slab of gas irradiated by the interstellar radiation field, either at constant density and temperature or in thermal and pressure equilibrium. The dominant thermal processes tend to form three distinct regions in the clouds. At low column densities cooling is dominated by ionised metals such as Si\textsc{ii}, Fe\textsc{ii}, Fe\textsc{iii} and C\textsc{ii}, which are balanced by photoheating, primarily from H\textsc{i}. Once the hydrogen-ionising radiation becomes attenuated by neutral hydrogen, photoelectric dust heating dominates, while C\textsc{ii} becomes dominant for cooling. Finally, dust shielding triggers the formation of CO and suppresses photoelectric heating. The dominant coolants in this fully shielded region are H$_{2}$ and CO. The column density of the H\textsc{i}-H$_{2}$ transition predicted by our model is lower at higher density (or at higher pressure for gas clouds in pressure equilibrium) and at higher metallicity, in agreement with previous PDR models. We also compare the H\textsc{i}-H$_{2}$ transition in our model to two prescriptions for molecular hydrogen formation that have been implemented in hydrodynamic simulations. 
\end{abstract}

\begin{keywords}
  astrochemistry - ISM: atoms - ISM: clouds - ISM: molecules - molecular processes - galaxies: ISM.
\end{keywords}

\section{Introduction}

The thermal evolution of gas is an important component of hydrodynamic simulations of galaxy formation as it determines how quickly the gas can cool and collapse to form dense structures, and ultimately stars. The star formation in such simulations can be limited to the cold phase of the interstellar medium (ISM) if we have sufficient resolution to resolve the Jeans mass ($M_{\rm{J}} \propto \rho^{-1/2} T^{3/2}$) in the cold gas. It is desirable to include a multi-phase treatment of the ISM, as this will produce a more realistic description of the distribution of star formation within the galaxy, along with the resulting impact of stellar feedback on the ISM and the galaxy as a whole \citep[e.g.][]{ceverino09,governato10,halle13,hopkins13}. Therefore, it is important that we correctly follow the thermal evolution of gas between the warm ($T \sim 10^{4}$ K) and cold ($T \la 100$ K) phases of the ISM in these simulations. 

The radiative cooling rate depends on the chemical abundances in the gas, including the ionisation balance, and hence on its chemical evolution. However, following the full non-equilibrium chemistry of the ISM within a hydrodynamic simulation can be computationally expensive, as it requires us to integrate a system of stiff differential equations that involves hundreds of species and thousands of reactions. Therefore, many existing cosmological hydrodynamic simulations use tabulated cooling rates assuming chemical (including ionisation) equilibrium. For example, the cosmological simulations that were run as part of the OverWhelmingly Large Simulations project \citep[OWLS;][]{schaye10} use the pre-computed cooling functions of \citet{wiersma09}, which were calculated using \textsc{Cloudy}\footnote{\url{http://nublado.org/}} \citep{ferland98,ferland13} as a function of temperature, density and abundances of individual elements assuming ionisation equilibrium in the presence of the \citet{haardt01} extragalactic UV background. However, this assumption of ionisation equilibrium may not remain valid if the cooling or dynamical time-scale becomes short compared to the chemical time-scale \citep[e.g.][]{kafatos73,gnat07,oppenheimer13a,vasiliev13} or if the UV radiation field is varying in time \citep[e.g.][]{oppenheimer13b}. 

In the first paper of this series (\citealt{richings14}; hereafter paper I) we presented a chemical network to follow the non-equilibrium thermal and chemical evolution of interstellar gas. Using this model we investigated the chemistry and cooling properties of optically thin interstellar gas in the ISM and identified the dominant coolants for gas exposed to various UV radiation fields. We also looked at the impact that non-equilibrium chemistry can have on the cooling rates and chemical abundances of such gas.

In this paper we extend our thermo-chemical model to account for gas that is shielded from the incident UV radiation field by some known column density. We focus on physical conditions with densities $10^{-2} \, \text{cm}^{-3} \la n_{\rm{H_{tot}}} \la 10^{4} \, \text{cm}^{-3}$ and temperatures of $10^{2} \, \text{K} \la T \la 10^{4} \, \text{K}$. This is most relevant to gas that is cooling from the warm phase to the cold phase of the ISM. We apply our model to a one-dimensional plane-parallel slab of gas that is irradiated by the \citet{black87} interstellar radiation field to investigate how the chemistry and cooling properties of the gas change as it becomes shielded from the UV radiation, both by dust and by the gas itself. The spectral shape of the radiation field will change with the depth into the cloud as high energy photons are able to penetrate deeper. Hence, the major coolants and heating processes will vary with column density. Such one-dimensional models are commonly used to model photodissociation regions (PDRs) and diffuse and dense clouds \citep[e.g.][]{tielens85,vandishoeck86,vandishoeck88,lepetit06,visser09,wolfire10}.

It has been suggested recently that the star formation rate of galaxies may be more strongly correlated to the molecular gas content than to atomic hydrogen \citep{wong02,schaye04,kennicutt07,leroy08,bigiel08,bigiel10}, although the more fundamental and physically relevant correlation may be with the cold gas content \citep{schaye04,krumholz11a,glover12}. Motivated by this link between molecular hydrogen and star formation, a number of studies have implemented simple methods to follow the abundance of H$_{2}$ in numerical simulations of galaxies \citep[e.g.][]{pelupessy06,gnedin09,mckee10,christensen12}. We compare the H$_{2}$ fractions predicted by some of these methods to those calculated using our model to investigate the physical processes that determine the H\textsc{i}-H$_{2}$ transition and to explore in which physical regimes these various prescriptions remain valid.

This paper is organised as follows. In section~\ref{modelSummary} we summarise the thermo-chemical model presented in paper I, and in section~\ref{shielding_section} we describe how the photochemical rates are attenuated by dust and gas. We look at the photoionisation rates in section~\ref{shieldIon_section}, the photodissociation of molecular species in section~\ref{shieldDissoc_section} and the photoheating rates in section~\ref{photoheat_section}. In section~\ref{shieldResultsSect} we apply this model to a one-dimensional plane-parallel slab of gas to investigate the chemistry and cooling properties of the gas as it becomes shielded from the UV radiation field, and we compare these results with \textsc{Cloudy}. In section~\ref{H2modelsSection} we compare the time-dependent molecular H$_{2}$ fractions predicted by our model with two prescriptions for H$_{2}$ formation taken from the literature that have been implemented in hydrodynamic simulations. Finally, we discuss our results and conclusions in section~\ref{conclusions}.

\section{Thermo-Chemical Model}\label{modelSummary}

In this section we give a brief overview of the chemical and thermal processes that are included in our model. These are described in more detail in paper I.

We follow the evolution of 157 chemical species, including 20 molecules (H$_{2}$, H$_{2}^{+}$, H$_{3}^{+}$, OH, H${_2}$O, C$_{2}$, O$_{2}$, HCO$^{+}$, CH, CH$_{2}$, CH$_{3}^{+}$, CO, CH$^{+}$, CH$_{2}^{+}$, OH$^{+}$, H$_{2}$O$^{+}$, H$_{3}$O$^{+}$, CO$^{+}$, HOC$^{+}$ and O$_{2}^{+}$) along with electrons and all ionisation states of the 11 elements that dominate the cooling rate (H, He, C, N, O, Ne, Si, Mg, S, Ca and Fe). The rate equations of these species and the equation for the temperature evolution give us a system of 158 differential equations that we integrate from the initial conditions with \textsc{Cvode} (a part of the \textsc{Sundials}\footnote{\url{https://computation.llnl.gov/casc/sundials/main.html}} suite of non-linear differential/algebraic equation solvers), using the backward difference formula method and Newton iteration.

\subsection{Chemistry}

Our chemical network contains 907 reactions, including:

\begin{description}

\item[\textit{Collisional reactions.}] We include the collisional ionisation, radiative and di-electronic recombination and charge transfer reactions of all ionisation states of the 11 elements in our network. The rate coefficients of these reactions were tabulated as a function of temperature by \citet{oppenheimer13a} using \textsc{Cloudy} (they use version 10.00, but they have since produced updated versions of these tables using version 13.01 of \textsc{Cloudy}\footnote{These updated tables are available on the website: \url{http://noneq.strw.leidenuniv.nl}}). We also include reactions for the formation and destruction of molecular hydrogen taken from \citet{glover07,glover08} and others, and the CO network from \citet{glover10} with some small modifications (see section 2.5 of paper I).

\item[\textit{Photochemical reactions.}] We compute the optically thin photoionisation rates using the grey approximation cross sections of each atom and ion species for a given UV spectrum, which we calculate using the frequency dependent cross sections from \citet{verner95} and \citet{verner96}. We also include Auger ionisation, where photoionisation of inner shell electrons can lead to the ejection of multiple electrons by a single photon. For these we use the electron vacancy distribution probabilities from \citet{kaastra93}. We assume that the optically thin photodissociation rate of molecular hydrogen scales linearly with the number density of photons in the energy band $12.24 \, \text{eV} < h \nu < 13.51 \, \text{eV}$, normalised to the photodissociation rate in the presence of the \citet{black87} interstellar radiation field calculated by \textsc{Cloudy} (see section 2.2.2 of paper I). For the remaining molecular species, we use the photoionisation and photodissociation rates given by \citet{vandishoeck06} and \citet{visser09} where available, or \citet{glover10} otherwise. In paper I we only considered optically thin gas. In section~\ref{shielding_section} we describe how we modify these optically thin rates for shielded gas. 

\item[\textit{Cosmic ray ionisation.}] The primary ionisation rate of H\textsc{i} due to cosmic rays, $\zeta_{\rm{HI}}$, is a free parameter in our model. We use a default value of $\zeta_{\rm{HI}} = 2.5 \times 10^{-17} \, \text{s}^{-1}$ \citep{williams98}, although recent observations suggest that this could be an order of magnitude larger \citep[e.g.][]{indriolo12}. In paper I we consider the impact that increasing or decreasing our default value of $\zeta_{\rm{HI}}$ by a factor of ten can have on the abundances in fully shielded gas (see figure 6 in paper I). 

The cosmic ray ionisation rates of the other species are then scaled linearly with $\zeta_{\rm{HI}}$. The ratio of the ionisation rate of each species with respect to H\textsc{i} is assumed to be equal to the ratio of these ionisation rates given in the \textsc{umist} database\footnote{\url{http://www.udfa.net}} \citep{leteuff00} where available. For species that do not appear in this database, we calculate their cosmic ray ionisation rate with respect to $\zeta_{\rm{HI}}$ using \citet{lotz67}, \citet{silk70} and \citet{langer78}. For the molecular species, we use the cosmic ray ionisation and dissociation rates from table B3 of \citet{glover10}, again scaled with $\zeta_{\rm{HI}}$.

\item[\textit{Dust grain reactions.}] The formation rate of molecular hydrogen on dust grains is calculated using equation 18 from \citet{cazaux02}. We take a constant dust temperature $T_{\rm{dust}} = 10$ K, which we find has a negligible impact on our results, and we assume that the abundance of dust scales linearly with metallicity. We also include a small number of recombination reactions on dust grains, taken from \citet{weingartner01a}.

\end{description}

\subsection{Thermal processes}

For a complete list of cooling and heating processes in our model, see table 1 of paper I. Below we summarise some of the cooling and heating mechanisms that are most important in the diffuse ISM.

\begin{description}

\item[\textit{Metal-line cooling.}] \citet{oppenheimer13a} tabulate the radiative cooling rates of all ionisation states of the 11 elements in our chemical network as a function of temperature, calculated using \textsc{Cloudy} (as for the rate coefficients, they use version 10.00, but they have since produced updated versions of these tables using version 13.01 of \textsc{Cloudy}$^{3}$). We use these cooling rates for most metal species in our model. However, these rates assume that the radiative cooling is dominated by electron-ion collisions. For a small number of species this assumption can break down at low temperatures ($T \la 10^{3}$ K). We therefore found it necessary to calculate the cooling rates of O\textsc{i} and C\textsc{i} as a function of temperature and of H\textsc{i}, H\textsc{ii} and electron densities, using the same method as \citet{glover07}. This enables us to follow the radiative cooling rates of these species in regimes that are dominated by collisions with H\textsc{i} or H\textsc{ii}, as well as when electron-ion collisions dominate. We also tabulated the cooling rates of C\textsc{ii}, N\textsc{ii}, Si\textsc{ii} and Fe\textsc{ii} as functions of temperature and electron density using version 7.1 of the \textsc{Chianti} database\footnote{\url{http://www.chiantidatabase.org/chianti.html}} \citep{dere97,landi13}. 

\item[\textit{H$_{2}$ rovibrational cooling.}] We use the H$_{2}$ cooling function from \citet{glover08}, assuming an ortho- to para- ratio of 3:1. This includes collisional excitation of the rovibrational levels of molecular hydrogen by H\textsc{i}, H\textsc{ii}, H$_{2}$, He\textsc{i} and electrons. 

\item[\textit{Photoheating.}] We calculate the average excess energy of ionising photons for each species, given the incident UV spectrum, using the frequency dependent cross sections from \citet{verner95} and \citet{verner96}. We then multiply these by the corresponding photoionisation rate to obtain the photoheating rate for each species. See section~\ref{photoheat_section} for a description of how we modify these optically thin photoheating rates for shielded gas. 

\item[\textit{Photoelectric heating.}] The absorption of UV photons by dust grains can release electrons, and the excess energy that is absorbed by the electrons is quickly thermalised, which heats the gas. We calculate the photoelectric heating rate from dust grains using equations 1 and 2 from \citet{wolfire95}. 

\end{description}

\section{Shielding processes}\label{shielding_section}

In paper I we considered only optically thin gas. However, for shielded gas we need to consider how both the intensity and the shape of the UV spectrum change as the radiation field becomes attenuated by both dust and the gas itself. The photochemical rates presented in paper I involve integrals over photon frequency, but these are expensive to compute. For the optically thin rates we use the grey approximation to obtain the average cross section of each species, weighted by the frequency dependent radiation field, so that these integrals do not need to be re-evaluated at every timestep in the chemistry solver. However, this approximation becomes invalid if the shape of the UV spectrum is no longer invariant. 

In the following sections we describe how we modify the optically thin rates of photoionisation, photodissociation and photoheating for gas that is shielded by a known column density. To implement these methods in a hydrodynamic simulation, we would need to estimate this column density for each gas particle/cell. This is typically done by assuming that shielding occurs locally over some characteristic length scale $L$, so that the column density $N_{i}$ of a gas cell with density $n_{i}$ can be estimated from local quantities as:

\begin{equation}
N_{i} = n_{i} L.
\end{equation}

If the macroscopic velocity gradient $dv/dr$  is large with respect to local variations in the thermal line broadening, for example in turbulent molecular clouds, we can use the Sobolev length \citep{sobolev57}, which gives the length scale over which the Doppler shift of a line due to the velocity gradient is equal to the thermal width of the line:

\begin{equation}\label{sobolevEqn}
L_{\rm{Sob}} = \frac{v_{\rm{th}}}{\lvert dv/dr \rvert},
\end{equation}
where $v_{\rm{th}}$ is the thermal velocity. This method is applicable to the shielding of individual lines, for example in the self-shielding of molecular hydrogen. \citet{gnedin09} use a Sobolev-like approximation to estimate the column density using the density gradient rather than the velocity gradient:

\begin{equation}
L_{\rm{Sob}, \rho} = \frac{\rho}{\lvert \nabla \rho \rvert}.
\end{equation}
By integrating column densities along random lines of sight in their cosmological simulations, in which they are able to resolve individual giant molecular complexes, \citet{gnedin09} confirm that this approximation reproduces the true column density with a scatter of a factor $\sim 2$ in the range $3 \times 10^{20} \, \text{cm}^{-2} < N_{\rm{HI}} + 2 N_{\rm{H_{2}}} < 3 \times 10^{23} \, \text{cm}^{-2}$ (see their figure 1). 

An alternative approach is to assume that shielding occurs locally on scales of the Jeans length, $L_{\rm{Jeans}}$ \citep{schaye01a,schaye01b,rahmati13}. For example, \citet{hopkins13} integrate the density out to $L_{\rm{Jeans}}$ for each particle to obtain its shielding column density, which they use to attenuate the UV background. 

\citet{wolcottgreen11} use these three methods to estimate the self-shielding of molecular hydrogen and compare them to their calculations of the radiative transfer of Lyman Werner radiation through simulated haloes at redshift $z \sim 10$. They find that using the Sobolev length (equation~\ref{sobolevEqn}) is the most accurate method based only on local properties, compared to their radiative transfer calculations. 

Our chemical model could also be coupled to a full 3D radiative transfer simulation, in which the shielding of the UV radiation is computed by the radiative transfer solver rather than using the methods that we describe below. However, to include the impact of shielding on the shape of the UV spectrum, we require multiple frequency bins, as the spectral shape is assumed to remain constant within individual frequency bins. Such a calculation is likely to be computationally expensive. In a future work we shall compare our methods described below, with different approximations for the column density, to radiative transfer calculations to investigate in which physical conditions these various approximations can be applied in hydrodynamic simulations.

\subsection{Photoionisation}\label{shieldIon_section}

\begin{figure}
\centering
\mbox{
	\includegraphics[width=84mm]{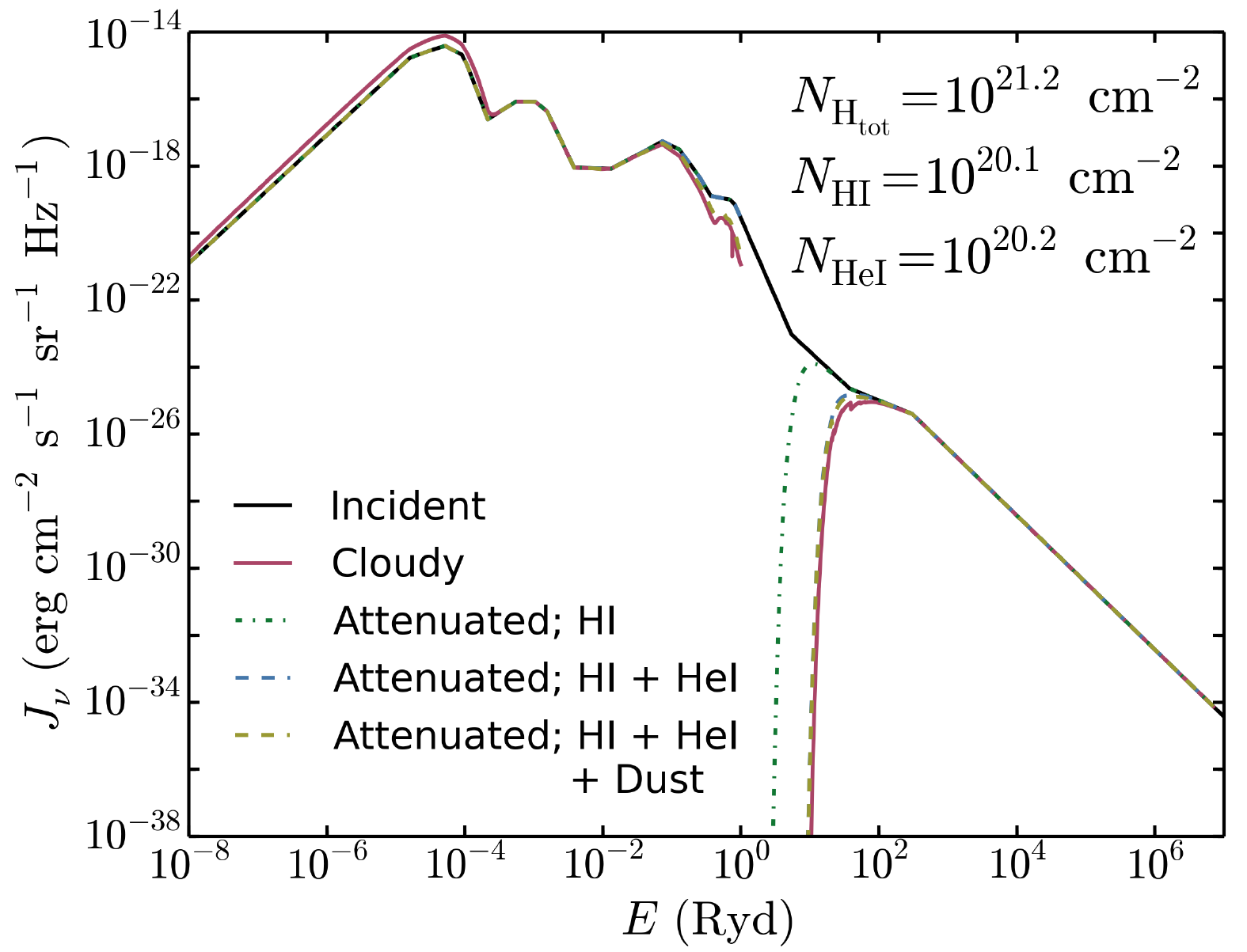}}
\caption{The effect of dust and gas shielding on the radiation spectrum. The incident spectrum of the \citet{black87} interstellar radiation field is shown by the black curve, and the coloured curves show the resulting spectrum after it has passed through a total hydrogen column density $N_{\rm{H_{tot}}} = 1.6 \times 10^{21} \, \text{cm}^{-2}$ at solar metallicity, as calculated by \textsc{Cloudy}, and the spectrum attenuated only by H\textsc{i}, by H\textsc{i} and He\textsc{i}, and by H\textsc{i}, He\textsc{i} and dust.}
\label{UVspectrumFig}
\end{figure}

The incident UV radiation field is attenuated by both dust and gas. The factor by which dust reduces the photoionisation rate of a species can be written as:

\begin{equation}\label{dust_shielding}
S_{\rm{d}}^{i} = \exp(- \gamma_{\rm{d}}^{i} A_{v}) = \exp \left(- 4.0 \times 10^{-22} \gamma_{\rm{d}}^{i} N_{\rm{H_{tot}}} Z / Z_{\odot} \right),
\end{equation}
where we can express this factor in terms of the visual extinction $A_{v}$ or the total intervening hydrogen column density $N_{\rm{H_{tot}}} = N_{\rm{HI}} + N_{\rm{HII}} + 2 N_{\rm{H_{2}}}$. Following \citet{krumholz11a}, we use $A_{v} / N_{\rm{H_{tot}}} = 4.0 \times 10^{-22} Z / Z_{\odot} \, \text{mag cm}^{2}$, which is intermediate between the values from the models of \citet{weingartner01b} for the Milky Way (with a visual extinction to reddening ratio $R_{v} = 3.1$ or $5.5$), the Large Magellanic Cloud and the Small Magellanic Cloud, and assumes that the dust content of the gas scales linearly with its metallicity. The factor $\gamma_{d}^{i}$ is a constant that is different for each species and depends on the range of photon energies that ionise that species. This factor accounts for the fact that the absorbing cross section of dust grains varies with photon energy, so the effective dust column density depends on which energy range we are interested in. We use the values for $\gamma_{d}^{i}$ calculated for the \citet{draine78} interstellar radiation field from table 3 of \citet{vandishoeck06} where available, or from table B2 of \citet{glover10} otherwise. Since Ca\textsc{i} and Ca\textsc{ii} were not included in either reference, we simply use the value of $\gamma_{d}^{i}$ from species that have similar ionisation energies (Mg\textsc{i} and C\textsc{i} respectively). 

While dust is the dominant source of absorption of UV radiation below 13.6 eV, above this limit UV radiation is strongly absorbed by neutral hydrogen, since photons at these energies are able to ionise hydrogen, which has a much higher cross sectional area $\sigma_{\rm{H}}$ ($\sim 10^{-17} \text{cm}^{2}$ at 1 Ryd) than the dust. Neutral helium can also significantly attenuate the UV radiation at energies above 24.6 eV. 

To illustrate how different components impact the spectrum at different energies, we show in figure~\ref{UVspectrumFig} the \citet{black87} interstellar radiation field and the resulting spectrum after it has passed through a total hydrogen column density $N_{\rm{H_{tot}}} = 1.6 \times 10^{21} \, \text{cm}^{-2}$ at solar metallicity, including an H\textsc{i} column density $N_{\rm{HI}} = 1.2 \times 10^{20} \, \text{cm}^{-2}$ and a He\textsc{i} column density $N_{\rm{HeI}} = 1.6 \times 10^{20} \, \text{cm}^{-2}$, as calculated by \textsc{Cloudy}. We then compare these to a spectrum attenuated by only H\textsc{i}, by H\textsc{i} and He\textsc{i}, and by H\textsc{i}, He\textsc{i} and dust, calculated using the frequency dependent cross sections of \citet{verner96} and the dust opacities from \citet{martin90}. We see that at energies just below 1 Ryd the UV radiation is suppressed by dust absorption, while the strong break above 1 Ryd is caused by neutral hydrogen. Comparing the green curve to the blue and yellow curves, we also see that absorption by neutral helium is important at higher energies, from a few Rydbergs up to 100 Ryd, while dust absorption is relatively unimportant at these higher energies. Note that the attenuated spectrum calculated by \textsc{Cloudy} (the red curve) includes thermal emission from dust grains, which is why it is slightly higher than the incident spectrum at low energies $E < 10^{-4}$ Ryd.

Throughout this paper we use the \citet{black87} interstellar radiation field, which consists of the background radiation field from IR to UV of \citet{mathis83}, the soft X-ray background from \citet{bregman85} and a blackbody with a temperature of 2.7 K. However, several of the photoionisation and photodissociation rates that we use for molecular species were calculated using the \citet{draine78} interstellar radiation field \citep[e.g.][]{vandishoeck06,visser09}. We normalise these rates by the radiation field strength in the energy band $6 \, \rm{eV} < h\nu < 13.6 \, \rm{eV}$, but the shape of these UV spectra are different. This will introduce an additional uncertainty in our results. 

While the column densities considered in the above example are typical of cold, atomic interstellar gas, there are certain regimes in which we also need to include the attenuation by additional species. For example, once the gas becomes molecular, the H$_{2}$ column density can be greater than the neutral hydrogen column density. Furthermore, at the high temperatures and low densities that are typical of the circumgalactic medium (for example $T \sim 10^{5}$ K, $n_{\rm{H_{tot}}} \sim 10^{-4} \, \text{cm}^{-3}$), helium is singly ionised and shielding of radiation above 54.4 eV by He\textsc{ii} can be important. Therefore, to calculate the shielded photoionisation rates of species with ionisation energies above 13.6 eV, we need to account for the attenuation by four species: H\textsc{i}, H$_{2}$, He\textsc{i} and He\textsc{ii}. For an incident spectrum with intensity $J_{\nu}$ per unit solid angle per unit frequency, the photoionisation rate of species $i$, $\Gamma_{i, \rm{thick}}$, is then given by:

\begin{align}\label{photoIonThick}
\Gamma_{i, \rm{thick}} = & \int_{\nu_{0, i}}^{\infty} \frac{4 \pi J_{\nu}}{h \nu} \exp(- N_{\rm{HI}} \sigma_{\nu, \rm{HI}} - N_{\rm{H_{2}}} \sigma_{\nu, \rm{H_{2}}} \notag \\
& - N_{\rm{HeI}} \sigma_{\nu, \rm{HeI}} - N_{\rm{HeII}} \sigma_{\nu, \rm{HeII}}) \sigma_{\nu, i} d\nu \, ,
\end{align}
where $\sigma_{\nu, i}$ and $\nu_{0, i}$ are respectively the frequency dependent cross section and the ionisation threshold frequency of species $i$.

Calculating these integrals at every timestep in the chemistry solver would be too computationally expensive, so instead we need to pre-compute them and tabulate the results. However, this would require us to tabulate the optically thick rates in four dimensions, one for each of the attenuating column densities, for every species. Such tables would require too much memory to be feasible. To reduce the size of the tables, we note that the H$_{2}$ cross section can be approximated as:

\begin{equation}\label{H2_approx_sigma}
\sigma_{\nu, \rm{H_{2}}} \approx \left\{
  \begin{array}{lll}
    3.00 \sigma_{\nu, \rm{HI}} & & h \nu > 15.4 \, \text{eV} = h \nu_{0, \rm{H_{2}}}\\
    0 & & \text{otherwise}.
  \end{array}
\right.
\end{equation}
Similarly, the He\textsc{ii} cross section can be approximated as:

\begin{equation}\label{HeII_approx_sigma}
\sigma_{\nu, \rm{HeII}} \approx \left\{
  \begin{array}{lll}
    0.75 \sigma_{\nu, \rm{HeI}} & & h \nu > 54.4 \, \text{eV} = h \nu_{0, \rm{HeII}} \\
    0 & & \text{otherwise}.
  \end{array}
\right.
\end{equation}

Using these approximations, we can divide the shielded photoionisation rates into three integrals as follows:

\begin{align}\label{photoIonThick_approx}
\Gamma_{i, \rm{thick}} & = \int_{\nu_{0, i}}^{\nu_{0, \rm{H_{2}}}} \frac{4 \pi J_{\nu}}{h \nu} \exp(- N_{\rm{HI}} \sigma_{\nu, \rm{HI}}) \sigma_{\nu, i} d\nu \, + \notag \\
& \int_{\nu_{0, \rm{H_{2}}}}^{\nu_{0, \rm{HeII}}} \frac{4 \pi J_{\nu}}{h \nu} \exp(- N_{\rm{H}}^{\rm{eff}} \sigma_{\nu, \rm{HI}} - N_{\rm{HeI}} \sigma_{\nu, \rm{HeI}}) \sigma_{\nu, i} d\nu \, + \notag \\ 
& \int_{\nu_{0, \rm{HeII}}}^{\infty} \frac{4 \pi J_{\nu}}{h \nu} \exp(- N_{\rm{H}}^{\rm{eff}} \sigma_{\nu, \rm{HI}} - N_{\rm{He}}^{\rm{eff}} \sigma_{\nu, \rm{HeI}}) \sigma_{\nu, i} d\nu \notag \\
& = \Gamma_{i, \rm{thin}} (S_{\rm{gas}, 1}^{i}(N_{\rm{HI}}) + S_{\rm{gas}, 2}^{i}(N_{\rm{H}}^{\rm{eff}}, N_{\rm{HeI}}) \notag \\
& + S_{\rm{gas}, 3}^{i}(N_{\rm{H}}^{\rm{eff}}, N_{\rm{He}}^{\rm{eff}})),
\end{align}
where the effective hydrogen and helium column densities are: 

\begin{align}
N_{\rm{H}}^{\rm{eff}} & = N_{\rm{HI}} + 3.00 N_{\rm{H_{2}}} \label{NH_eff_eqn} \\
N_{\rm{He}}^{\rm{eff}} & = N_{\rm{HeI}} + 0.75 N_{\rm{HeII}}. \label{NHe_eff_eqn}
\end{align}
These effective column densities also account for the attenuation by H$_{2}$ and He\textsc{ii} respectively. Note that equation~\ref{photoIonThick_approx} is valid for species with an ionisation threshold frequency $\nu_{0, i} < \nu_{0, \rm{H_{2}}}$. If $\nu_{0, i} \geq \nu_{0, \rm{H_{2}}}$, the first integral will be zero. Similarly, the second integral in equation~\ref{photoIonThick_approx} will also be zero if $\nu_{0, i} \geq \nu_{0, \rm{HeII}}$.

By using these approximations, we only need to create tables of $S_{\rm{gas}, \{1,2,3\}}$ in up to two dimensions, which greatly reduces the memory that they require. This approach is not exact, but we find that it introduces errors in $\Gamma_{i, \rm{thick}}$ of at most a few tens of per cent, and typically much less than this. In appendix~\ref{shielding_approximations} we show the relative errors in the photoionisation rates of each species that are introduced by these approximations. For each species $i$ we use equation~\ref{photoIonThick_approx} to tabulate the integrals $S_{\rm{gas}, 1}^{i}(N_{\rm{HI}})$, $S_{\rm{gas}, 2}^{i}(N_{\rm{H}}^{\rm{eff}}, N_{\rm{HeI}})$ and $S_{\rm{gas}, 3}^{i}(N_{\rm{H}}^{\rm{eff}}, N_{\rm{He}}^{\rm{eff}})$ as a function of the given column densities from 10$^{15}$ to 10$^{24}$ cm$^{-2}$ in intervals of 0.1 dex. Added together, these integrals give the ratio of the optically thick to the optically thin photoionisation rate of species $i$, $S_{\rm{gas}}^{i}(N_{\rm{HI}}, N_{\rm{H_{2}}}, N_{\rm{HeI}}, N_{\rm{HeII}})$.

\subsection{Photodissociation}\label{shieldDissoc_section}

The photodissociation rates of molecular species are attenuated by dust according to equation~\ref{dust_shielding}, where we take the values of $\gamma_{\rm{d}}^{i}$ calculated for the \citet{draine78} interstellar radiation field from table 2 of \citet{vandishoeck06} where available, or from table B2 of \citet{glover10} otherwise. In addition to dust shielding, the absorption of Lyman Werner radiation (i.e. photon energies $11.2 \, \text{eV} < h \nu < 13.6 \, \text{eV}$) by molecular hydrogen allows H$_{2}$ to become self-shielded. An accurate treatment of this effect would require us to solve the radiative transfer of the Lyman Werner radiation and to follow the level populations of the rovibrational states of the H$_{2}$ molecule, which would be computationally expensive. However, we can approximate the self-shielding of H$_{2}$ as a function of H$_{2}$ column density. For example, the following analytic fitting function from \citet{draine96} is commonly used in hydrodynamic simulations of galaxies and molecular clouds \citep[e.g.][]{glover07,glover10,gnedin09,christensen12,krumholz12}:

\begin{equation}\label{H2self_eqn}
S_{\rm{self}}^{\rm{H_{2}}} = \frac{1 - \omega_{\rm{H_{2}}}}{(1 + x / b_{5})^{\alpha}} + \frac{\omega_{\rm{H_{2}}}}{(1 + x)^{1/2}} \exp(-8.5 \times 10^{-4} (1 + x)^{1/2}),
\end{equation}
where $x \equiv N_{\rm{H_{2}}} / (5 \times 10^{14} \text{cm}^{-2})$, $\omega_{\rm{H_{2}}}$ and $\alpha$ are adjustable parameters (\citealt{draine96} use $\omega_{\rm{H_{2}}} = 0.035$ and $\alpha = 2$), $b_{5} \equiv b / (10^{5} \text{cm s}^{-1})$ and $b$ is the Doppler broadening parameter. This function was introduced by \citet{draine96} and was motivated by their detailed radiative transfer and photodissociation calculations. The suppression factor $S_{\rm{self}}^{\rm{H_{2}}}$ initially declines rapidly (indicating increased shielding) with $N_{\rm{H_{2}}}$ as individual Lyman Werner lines shield themselves. However, once these lines become saturated, the shielding is determined by the wings of the lines, leading to a shallow power law dependence $\sim N_{\rm{H_{2}}}^{-1/2}$. Finally, at high column densities the lines overlap, creating an exponential cut off in the self-shielding function.

Some authors have used equation~\ref{H2self_eqn} with different values for some of the parameters. For example, \citet{gnedin09} and \citet{christensen12} adopt a value $\omega_{\rm{H_{2}}} = 0.2$, as this gives better agreement between their model for H$_{2}$ formation and observations of atomic and molecular gas fractions in nearby galaxies. 

\citet{wolcottgreen11} also investigate the validity of equation~\ref{H2self_eqn} and compare it to their detailed radiative transfer calculations of Lyman Werner radiation through simulated haloes at high redshift ($z \sim 10$). They find that equation~\ref{H2self_eqn}, with $\omega_{\rm{H_{2}}} = 0.035$ and $\alpha = 2$, as used by \citet{draine96}, underestimates the value of $S_{\rm{self}}^{\rm{H_{2}}}$ (i.e. it predicts too strong self-shielding) by up to an order of magnitude in warm gas (with $T \ga 500$ K). They argue that this discrepancy arises because the assumptions made in \citet{draine96} are only accurate for cold gas in which only the lowest rotational states of H$_{2}$ are populated. However, they obtain better agreement with their calculations (within $\sim 15$ per cent) if they use equation~\ref{H2self_eqn} with $\alpha = 1.1$.

We have compared the H$_{2}$ self-shielding function of \citet{draine96} and the modification to this function suggested by \citet{wolcottgreen11}, with $\alpha = 1.1$, to the ratio of the optically thick to optically thin H$_{2}$ photodissociation rates predicted by \textsc{Cloudy} in primordial gas. Details of this comparison can be found in appendix~\ref{H2self_comparison_section}. We find that neither function produces satisfactory agreement with \textsc{Cloudy}. For example, both overestimate $S_{\rm{self}}^{\rm{H_{2}}}$ compared to \textsc{Cloudy} by a factor $\sim 3$ at H$_{2}$ column densities $N_{\rm{H_{2}}} \ga 10^{17} \, \rm{cm}^{-2}$ in gas with a temperature $T = 100$ K (see figure~\ref{H2self_fig}). We also find that the temperature dependence of $S_{\rm{self}}^{\rm{H_{2}}}$ predicted by \textsc{Cloudy} is not accurately reproduced by the modified self-shielding function from \citet{wolcottgreen11}. Furthermore, \citet{wolcottgreen11} only consider gas in which the Doppler broadening is purely thermal. However, if it is dominated by turbulence then equation~\ref{H2self_eqn} will be independent of temperature. 

To obtain a better fit to the H$_{2}$ self-shielding predicted by \textsc{Cloudy} (with its `big H2' model), we modified equation~\ref{H2self_eqn} as follows:

\begin{align}\label{H2self_mod_eqn}
S_{\rm{self}}^{\rm{H_{2}}} = & \frac{1 - \omega_{\rm{H_{2}}}(T)}{(1 + x^{\prime} / b_{5})^{\alpha(T)}} \exp(-5 \times 10^{-7} (1 + x^{\prime})) + \notag \\ 
& \frac{\omega_{\rm{H_{2}}}(T)}{(1 + x^{\prime})^{1/2}} \exp(-8.5 \times 10^{-4} (1 + x^{\prime})^{1/2}),
\end{align}
where $x^{\prime} = N_{\rm{H_{2}}} / N_{\rm{crit}}(T)$. To reproduce the temperature dependence of $S_{\rm{self}}^{\rm{H_{2}}}$ that we see in \textsc{Cloudy}, we fit the parameters $\omega_{\rm{H_{2}}}(T)$, $\alpha(T)$ and $N_{\rm{crit}}(T)$ as functions of the temperature $T$. We obtain the best agreement with \textsc{Cloudy} using:

\begin{equation}\label{omegaEqn}
\omega_{\rm{H_{2}}}(T) = 0.013 \left[ 1 + \left( \frac{T}{2700 \, \rm{K}} \right)^{1.3} \right]^{\frac{1}{1.3}} \! \! \exp\left[ - \left( \frac{T}{3900 \, \rm{K}} \right)^{14.6} \right], 
\end{equation}

\begin{equation}\label{alphaEqn}
\alpha(T) =  \left\{
  \begin{array}{lll}
    1.4 & & T < 3000 \, \rm{K} \\
    \left( \frac{T}{4500 \, \rm{K}} \right)^{-0.8} & & 3000 \leq T < 4000 \, \rm{K} \\
    1.1 & & T \geq 4000 \, \rm{K},
  \end{array}
\right.
\end{equation}

\begin{equation}\label{NcritEqn}
\frac{N_{\rm{crit}}(T)}{10^{14} \rm {cm}^{-2}} =  \left\{
  \begin{array}{lll}
    1.3 \left[ 1 + \left(\frac{T}{600 \, \rm{K}} \right)^{0.8} \right] & & T < 3000 \, \rm{K} \\
    \left( \frac{T}{4760 \, \rm{K}} \right)^{-3.8} & & 3000 \leq T < 4000 \, \rm{K} \\
    2.0 & & T \geq 4000 \, \rm{K}.
  \end{array}
\right.
\end{equation}

The H$_{2}$ self-shielding factor that we obtain with equations~\ref{H2self_mod_eqn} to \ref{NcritEqn} agrees with \textsc{Cloudy} to within 30 per cent at 100 K for $N_{\rm{H_{2}}} < 10^{21} \, \rm{cm}^{-2}$, and to within 60 per cent at 5000 K for $N_{\rm{H_{2}}} < 10^{20} \, \rm{cm}^{-2}$ (see appendix~\ref{H2self_comparison_section}).

Doppler broadening of the Lyman Werner lines suppresses self-shielding. In a hydrodynamic simulation there are a number of ways we can estimate $b$ (see \citet{glover07} for a more detailed discussion of some of the approximations that have been used in the literature). In this paper we shall include turbulence with a constant velocity dispersion of 5 km s$^{-1}$ (unless stated otherwise), which corresponds to a turbulent Doppler broadening parameter of $b_{\rm{turb}} = 7.1 \, \rm{km} \, \rm{s}^{-1}$, as used by \citet{krumholz12}. We also include thermal Doppler broadening $b_{\rm{therm}}$, which is related to the temperature $T$ by:

\begin{equation}
b_{\rm{therm}} = \sqrt{\frac{2 k_{B} T}{m_{\rm{H_{2}}}}}, 
\end{equation}
where $m_{\rm{H_{2}}}$ is the mass of an H$_{2}$ molecule. The total Doppler broadening parameter is then:

\begin{equation}
b = \sqrt{b_{\rm{therm}}^{2} + b_{\rm{turb}}^{2}}. 
\end{equation}

CO has a dissociation energy of 11.1 eV, which is very close to the lower energy of the Lyman Werner band. Therefore, photons in the Lyman Werner band are also able to dissociate CO. CO is photodissociated via absorptions in discrete lines (dissociation via continuum absorption is negligible for CO), so it can become self-shielded once these lines are saturated, but it can also be shielded by H$_{2}$, as its lines also lie in the Lyman Werner band. The shielding factor of CO due to H$_{2}$ and CO ($S_{\rm{self, H_{2}}}^{\rm{CO}}$) has been tabulated in two dimensions as a function of $N_{\rm{H_{2}}}$ and $N_{\rm{CO}}$ by \citet{visser09} for different values of the Doppler broadening and excitation temperatures of CO and H$_{2}$. High resolution versions of these tables can be found on their website\footnote{\url{home.strw.leidenuniv.nl/~ewine/photo}}. We use their table calculated for Doppler widths of CO and H$_{2}$ of $0.3 \, \rm{km} \, \rm{s}^{-1}$ and $3.0 \, \rm{km} \, \rm{s}^{-1}$ respectively, and excitation temperatures of CO and H$_{2}$ of 50.0 K and 353.6 K respectively, with the elemental isotope ratios of Carbon and Oxygen from \citet{wilson99} for the local ISM. 

Additionally, dust can shield CO, where the dust shielding factor is given by equation~\ref{dust_shielding} (with $\gamma_{\rm{d}}^{\rm{CO}} = 3.53$). 

To summarise, the optically thick photodissociation rates of H$_{2}$ and CO are:

\begin{align}
\Gamma_{\rm{H_{2}, thick}} &= \Gamma_{\rm{H_{2}, thin}} S_{\rm{d}}^{\rm{H_{2}}}(N_{\rm{H_{tot}}}, Z) S_{\rm{self}}^{\rm{H_{2}}}(N_{\rm{H_{2}}}) \\
\Gamma_{\rm{CO, thick}} &= \Gamma_{\rm{CO, thin}} S_{\rm{d}}^{\rm{CO}}(N_{\rm{H_{tot}}}, Z) S_{\rm{self, H_{2}}}^{\rm{CO}}(N_{\rm{CO}}, N_{\rm{H_{2}}}),
\end{align}
while the photoionisation and photodissociation rates of the remaining species are:

\begin{equation}
\Gamma_{i, \rm{thick}} = \Gamma_{i} S^{i},
\end{equation}
where $ S^{i} = S_{\rm{d}}^{i}(N_{\rm{H_{tot}}}, Z)$ if the ionisation energy is below 13.6 eV, or $S_{\rm{gas}}^{i}(N_{\rm{HI}}, N_{\rm{H_{2}}}, N_{\rm{HeI}}, N_{\rm{HeII}})$ otherwise.

To calculate the optically thick rates, we thus need to specify the column densities of H\textsc{i}, He\textsc{i}, He\textsc{ii}, H$_{2}$, H$_{\rm{tot}}$ and CO, and the metallicity. 

\subsection{Photoheating}\label{photoheat_section}

The photoheating rate of a species is the photoionisation rate $\Gamma_{i}$ multiplied by the average excess energy of the ionising photons $\left< \epsilon_{i} \right>$ (see equations 3.5 and 3.6 from paper I). As the gas becomes shielded, the shape of the UV spectrum changes, as more energetic photons are able to penetrate deeper into the gas, thus $\left< \epsilon_{i} \right>$ increases. In particular, species that have an ionisation energy above 13.6 eV may be strongly affected by absorption by H\textsc{i}, H$_{2}$, He\textsc{i} and He\textsc{ii} (see section~\ref{shieldIon_section}). For UV radiation attenuated by column densities $N_{\rm{HI}}$, $N_{\rm{H_{2}}}$, $N_{\rm{HeI}}$ and $N_{\rm{HeII}}$, the average excess energy of ionising photons for species $i$ is:

\begin{align}\label{epsilonThickEqnExact}
\left< \right. & \left. \!\! \epsilon_{i, \rm{thick}} \right> = \notag \\
 & \left[ \int_{\nu_{0, i}}^{\infty} \frac{4 \pi J_{\nu}}{h \nu} \exp(- N_{\rm{HI}} \sigma_{\nu, \rm{HI}} - N_{\rm{H_{2}}} \sigma_{\nu, \rm{H_{2}}} - N_{\rm{HeI}} \sigma_{\nu, \rm{HeI}} \right. \notag \\
& \left. \vphantom{\int} - N_{\rm{HeII}} \sigma_{\nu, \rm{HeII}}) (h \nu - h \nu_{0, i}) \sigma_{\nu, i} d\nu \right] \left. \vphantom{\int} \middle/ \right. \notag \\
& \left[ \int_{\nu_{0, i}}^{\infty} \frac{4 \pi J_{\nu}}{h \nu} \exp(- N_{\rm{HI}} \sigma_{\nu, \rm{HI}} - N_{\rm{H_{2}}} \sigma_{\nu, \rm{H_{2}}} - N_{\rm{HeI}} \sigma_{\nu, \rm{HeI}} \right. \notag \\
& \left. \vphantom{\int}- N_{\rm{HeII}} \sigma_{\nu, \rm{HeII}}) \sigma_{\nu, i} d\nu \right].
\end{align}

To pre-compute these integrals, we would require four dimensional tables. However, as described in section~\ref{shieldIon_section}, we can use the approximations in equations~\ref{H2_approx_sigma} and \ref{HeII_approx_sigma} to reduce the size of these tables. With these approximations, equation~\ref{epsilonThickEqnExact} becomes:

\begin{align}\label{epsilonThickEqn}
\left< \right. & \left. \!\! \epsilon_{i, \rm{thick}} \right> = \notag \\
 & \left[ \int_{\nu_{0, i}}^{\nu_{0, \rm{H_{2}}}} \frac{4 \pi J_{\nu}}{h \nu} \exp(- N_{\rm{HI}} \sigma_{\nu, \rm{HI}}) (h \nu - h \nu_{0, i}) \sigma_{\nu, i} d\nu \, + \right. \notag \\
 & \left. \int_{\nu_{0, \rm{H_{2}}}}^{\nu_{0, \rm{HeII}}} \frac{4 \pi J_{\nu}}{h \nu} \exp(- N_{\rm{H}}^{\rm{eff}} \sigma_{\nu, \rm{HI}} \right. \notag \\
 & \left. - N_{\rm{HeI}} \sigma_{\nu, \rm{HeI}}) (h \nu - h \nu_{0, i}) \sigma_{\nu, i} d\nu \, + \right. \notag \\
 & \left. \int_{\nu_{0, \rm{HeII}}}^{\infty} \frac{4 \pi J_{\nu}}{h \nu} \exp(- N_{\rm{H}}^{\rm{eff}} \sigma_{\nu, \rm{HI}} \right. \notag \\
 & \left. - N_{\rm{He}}^{\rm{eff}} \sigma_{\nu, \rm{HeI}}) (h \nu - h \nu_{0, i}) \sigma_{\nu, i} d\nu \right] \left. \vphantom{\int} \middle/ \right. \notag \\
 & \left[ \int_{\nu_{0, i}}^{\nu_{0, \rm{H_{2}}}} \frac{4 \pi J_{\nu}}{h \nu} \exp(- N_{\rm{HI}} \sigma_{\nu, \rm{HI}}) \sigma_{\nu, i} d\nu \, + \right. \notag \\
 & \left. \int_{\nu_{0, \rm{H_{2}}}}^{\nu_{0, \rm{HeII}}} \frac{4 \pi J_{\nu}}{h \nu} \exp(- N_{\rm{H}}^{\rm{eff}} \sigma_{\nu, \rm{HI}} - N_{\rm{HeI}} \sigma_{\nu, \rm{HeI}}) \sigma_{\nu, i} d\nu \, + \right. \notag \\
 & \left. \int_{\nu_{0, \rm{HeII}}}^{\infty} \frac{4 \pi J_{\nu}}{h \nu} \exp(- N_{\rm{H}}^{\rm{eff}} \sigma_{\nu, \rm{HI}} - N_{\rm{He}}^{\rm{eff}} \sigma_{\nu, \rm{HeI}}) \sigma_{\nu, i} d\nu \right],
\end{align}
where the effective hydrogen and helium column densities, $N_{\rm{H}}^{\rm{eff}}$ and $N_{\rm{He}}^{\rm{eff}}$, are given in equations~\ref{NH_eff_eqn} and \ref{NHe_eff_eqn}, and account for the attenuation by H$_{2}$ and He\textsc{ii} respectively.

For each species with an ionisation energy above 13.6 eV, we tabulate the six integrals in equation~\ref{epsilonThickEqn} as a function of the given column densities from 10$^{15}$ to 10$^{24}$ cm$^{-2}$ in intervals of 0.1 dex.

\section{Chemistry and cooling in shielded gas}\label{shieldResultsSect}

In this section we look at gas that is illuminated by a radiation field that is attenuated by some column density. We consider a one-dimensional plane-parallel slab of gas with solar metallicity that is illuminated from one side by the \citet{black87} interstellar radiation field. Throughout this paper we use the default solar abundances assumed by \textsc{Cloudy}, version 13.01 (see for example table 1 in \citealt{wiersma09}). In particular, we take the solar metallicity to be $Z_{\odot} = 0.0129$. The slab is divided into cells such that the total hydrogen column density from the illuminated face of the slab increases from $10^{14} \, \text{cm}^{-2}$ to $10^{24} \, \text{cm}^{-2}$ in increments of 0.01 dex. Using the methods described in section~\ref{shielding_section}, we then use the resulting column densities to calculate the attenuated photoionisation, photodissociation and photoheating rates and hence to solve for the chemical and thermal evolution in each cell.

The thermo-chemical evolution of a cell will depend on the chemical state of all cells between it and the illuminated face of the slab, since H\textsc{i}, He\textsc{i}, He\textsc{ii}, H$_{2}$ and CO contribute to the shielding of certain species. We therefore integrate the thermo-chemistry over a timestep that is determined such that the relative change in the column densities of H\textsc{i}, He\textsc{i}, He\textsc{ii}, H$_{2}$ and CO in each cell will be below some tolerance $\epsilon$ (which we take to be 0.1), as estimated based on their change over the previous timestep. In other words, the timestep $\Delta t$ is given by:

\begin{equation}
\Delta t = \min \left( \frac{\epsilon N_{i} \Delta t^{\rm{prev}}}{\Delta N_{i}^{\rm{prev}} + \psi} \right), \\
\end{equation}
where $N_{i}$ is the column density of species $i$ and $\Delta N_{i}^{\rm{prev}}$ is the change in $N_{i}$ over the previous timestep $\Delta t^{\rm{prev}}$. We take the minimum over the five species that contribute to shielding and over all gas cells. $\psi$ is a small number that is introduced to prevent division by zero - we take $\psi = 10^{-40}$. At the end of each timestep we then update the column densities of these five species for each cell.

\subsection{Comparison with \textsc{Cloudy}}

\begin{figure*}
\centering
\mbox{
	\includegraphics[width=148mm]{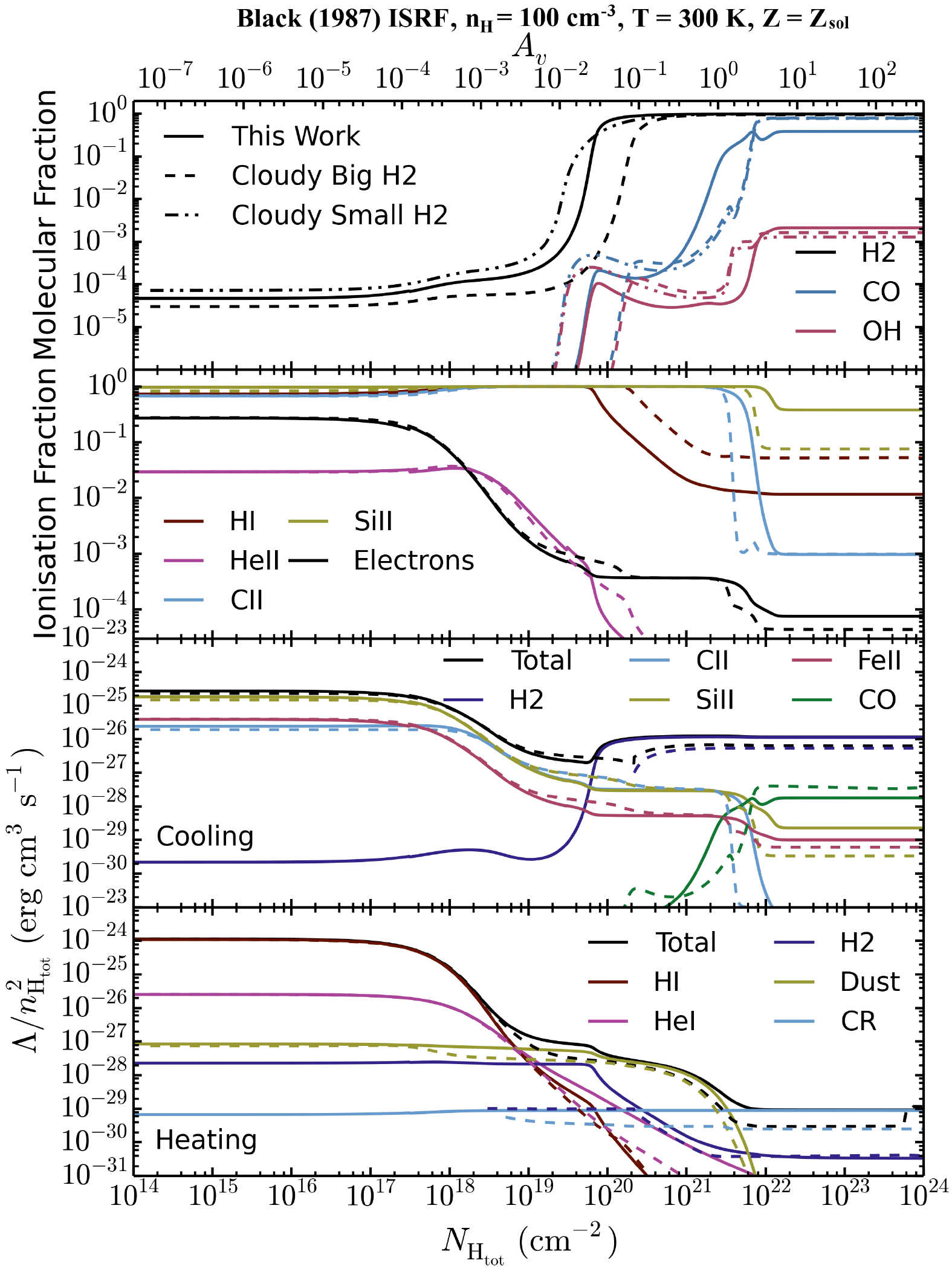}}
\caption{Chemistry and cooling properties of a one-dimensional plane-parallel slab of gas illuminated by the \citet{black87} interstellar radiation field, plotted as a function of the total hydrogen column density into the gas cloud, assuming solar metallicity with constant density $n_{\rm{H}} = 100 \, \text{cm}^{-3}$ and constant temperature $T = 300$ K. The corresponding dust extinction is marked on the top x-axis, where we have used $A_{v} / N_{\rm{H_{tot}}} = 4 \times 10^{-22} \, \text{mag cm}^{2}$ for a solar metallicity gas (see section~\ref{shieldIon_section}). We compare our model \textit{(solid lines)} to \textsc{Cloudy} using its big H2 model \textit{(dashed lines)} and its small H2 model \textit{(dot-dot-dashed lines; top panel only)} in chemical equilibrium. \textbf{Top panel:} equilibrium molecular fractions ($2 n_{\rm{H_{2}}} / n_{\rm{H_{tot}}}$, $n_{\rm{CO}} / n_{\rm{C_{tot}}}$ and $n_{\rm{OH}} / n_{\rm{O_{tot}}}$); \textbf{second panel:} equilibrium ionisation fractions; \textbf{third panel:} equilibrium cooling rates; \textbf{bottom panel:} equilibrium heating rates. The small discontinuity in the electron abundance at $N_{\rm{H_{tot}}} \sim 3 \times 10^{17} \, \text{cm}^{-2}$ is due to the recombination of hydrogen, which switches from case A to case B when the H\textsc{i} optical depth is unity.}
\label{shieldedCombinedFig}
\end{figure*}  

We use \textsc{Cloudy} version 13.01 to calculate the abundances and the cooling and heating rates in chemical equilibrium as a function of the total hydrogen column density, $N_{\rm{H_{tot}}}$, and compare them to the results from our model. These were calculated for a number of temperatures and densities, which were held fixed across the entire gas slab for this comparison. We add turbulence to the \textsc{Cloudy} models with a Doppler broadening parameter $b_{\rm{turb}} = 7.1 \, \text{km} \, \text{s}^{-1}$, in agreement with the default value that we use. 

\textsc{Cloudy} has two possible models for the microphysics of molecular hydrogen. Their `big H2' model follows the level populations of 1893 rovibrational states of molecular hydrogen, and the photodissociation rates from the various electronic and rovibrational transitions via the Solomon process are calculated self-consistently. However, as this is computationally expensive, \textsc{Cloudy} also has a `small H2' model in which the ground electronic state of molecular hydrogen is split between two vibrational states, a ground state and a single vibrationally excited state, following the approach of \citet{tielens85}. These two states are then treated as separate species in the chemical network. The photodissociation rates of H$_{2}$ in the small H2 model are taken from \citet{elwert05} by default (although there are options to use alternative dissociation rates). We primarily focus on the big H2 model for this comparison, as it includes a more complete treatment of the microphysics involved, although we also show the molecular abundances from the small H2 model to illustrate the differences between these two \textsc{Cloudy} models.

The comparison with \textsc{Cloudy} is shown in figure~\ref{shieldedCombinedFig} for gas at solar metallicity with a constant density $n_{\rm{H}} = 100 \, \text{cm}^{-3}$ and a temperature $T = 300$ K. More examples at other densities and temperatures can be found on our website\footnote{\url{http://noneqism.strw.leidenuniv.nl}}. In the top two panels we compare some of the equilibrium molecular and ionisation fractions. Compared to the big H2 model in \textsc{Cloudy}, the H\textsc{i}-H$_{2}$ transition occurs at a somewhat lower column density in our model, with a molecular hydrogen fraction $x_{\rm{H_{2}}} = 0.5$ at $N_{\rm{H_{tot}}} \approx 8.1 \times 10^{19} \, \text{cm}^{-2}$, compared to $N_{\rm{H_{tot}}} \approx 2.8 \times 10^{20} \, \text{cm}^{-2}$ in \textsc{Cloudy}. Below this transition the molecular hydrogen fraction tends towards a value $x_{\rm{H_{2}}} \approx 5 \times 10^{-5}$ in our model. This fraction is sufficiently high for the H$_{2}$ to shield itself from the Lyman Werner radiation before dust shielding becomes important (with $S_{\rm{d}}^{\rm{H_{2}}} \sim 0.1$ at $A_{v} \approx 0.6$), as the self-shielding is significant at relatively low H$_{2}$ column densities (with $S_{\rm{self}}^{\rm{H_{2}}} \sim 0.1$ at $N_{\rm{H_{2}}} \approx 6 \times 10^{15} \, \text{cm}^{-2}$, as defined in equation~\ref{H2self_mod_eqn}). The H$_{2}$ fraction in the photodissociated region is $x_{\rm{H_{2}}} \approx 3 \times 10^{-5}$ for \textsc{Cloudy}'s big H2 model, which is slightly lower than in our model. This explains why the gas becomes self-shielded at a somewhat higher total column density in \textsc{Cloudy} than in our model. This discrepancy occurs because we use a different photodissociation rate. As discussed in paper I, the dissociation rate of H$_{2}$ via the Solomon process depends on the level populations of the rovibrational states of H$_{2}$. This is calculated self-consistently in the big H2 model of \textsc{Cloudy}, whereas we must use an approximation to the photodissociation rate, as we do not follow the rovibrational levels of H$_{2}$. We therefore miss, for example, the dependence of the photodissociation rate on density, which affects the rovibrational level populations. 

For comparison, we also show the molecular hydrogen abundance predicted by the small H2 model of \textsc{Cloudy} in the top panel of figure~\ref{shieldedCombinedFig} \textit{(black dot-dot-dashed line)}. The H\textsc{i}-H$_{2}$ transition in the small H2 model is closer to our model than the big H2 model, with $x_{\rm{H_{2}}} = 0.5$ at $N_{\rm{H_{tot}}} \approx 1.1 \times 10^{20} \, \rm{cm}^{-2}$. However, the transition in our model is somewhat steeper than in \textsc{Cloudy}'s small H2 model. 

While the H\textsc{i}-H$_{2}$ transition in this example, with constant temperature and density, is caused by H$_{2}$ self-shielding, we would not expect temperatures as low as 300 K in the photodissociated region, nor would we expect densities as high as 100 cm$^{-3}$. This example therefore overestimates the importance of H$_{2}$ self-shielding. In section~\ref{isobarShielding} we address this issue by considering a cloud that is in thermal and pressure equilibrium.

After the H\textsc{i}-H$_{2}$ transition the neutral hydrogen abundance $x_{\rm{HI}}$ tends to $x_{\rm{HI}} \approx 0.01$ at $N_{\rm{H_{tot}}} \ga 10^{22} \, \text{cm}^{-2}$ in our model, compared to $x_{\rm{HI}} \approx 0.05$ in \textsc{Cloudy}. This difference arises because the cosmic ray dissociation rate af H$_{2}$ that we use in our model, which is based on the rates in the \textsc{umist} database, is an order of magnitude lower than that used in \textsc{Cloudy}.

The onset of the H\textsc{i}-H$_{2}$ transition leads to a rise in the CO abundance, although most carbon is still in C\textsc{ii} at this transition. At higher column densities the dust is able to shield the photodissociation of CO, which becomes fully shielded at $N_{\rm{H_{tot}}} \ga 10^{22} \, \rm{cm}^{-2}$. This second transition also corresponds to a drop in the abundances of singly ionised metals with ionisation energies lower than 1 Ryd, such as C\textsc{ii} and Si\textsc{ii}, whose photoionisation rates are attenuated by dust rather than by neutral hydrogen and helium. These species become shielded at a slightly higher column density in our model than in \textsc{Cloudy} (for example, C\textsc{ii} has an ionisation fraction $x_{\rm{CII}} = 0.1$ at $N_{\rm{H_{tot}}} \approx 7 \times 10^{21} \, \text{cm}^{-2}$ in our model, compared to $N_{\rm{H_{tot}}} \approx 4 \times 10^{21} \, \text{cm}^{-2}$ in \textsc{Cloudy}). 

Previous models of photodissociation regions have demonstrated that molecules such as H$_{2}$O will freeze out at large depths, with visual extinction $A_{v} \ga 10$ \citep[e.g.][]{hollenbach09,hollenbach12}. However, we do not include the freeze out of molecules in our model (and we also exclude molecule freeze out in \textsc{Cloudy} for comparison with our model). This will affect the gas phase chemistry, so the abundances predicted by our model in figure~\ref{shieldedCombinedFig} are likely to be unrealistic at the highest depths shown here ($A_{v} \ga 10$). 

In the bottom two panels of figure~\ref{shieldedCombinedFig} we show the total cooling and heating rates for this example, along with the contributions from selected individual species. Before the H\textsc{i}-H$_{2}$ transition the cooling is dominated by Si\textsc{ii}, Fe\textsc{ii} and C\textsc{ii}, with heating coming primarily from photoionisation of neutral hydrogen. After this transition the photoheating rates drop rapidly, with the main heating mechanism being photoelectric dust heating, while the total cooling rates are lower and are primarily from molecular hydrogen. Once dust shielding begins to significantly attenuate the UV radiation field below 13.6 eV at a column density $N_{\rm{H_{tot}}} \sim 10^{21} \, \text{cm}^{-2}$, the photoelectric heating rate falls sharply. For $N_{\rm{H_{tot}}} \ga 10^{22} \, \text{cm}^{-2}$ the heating rates are dominated by cosmic rays. In our model this is primarily from cosmic ray ionisation heating of H$_{2}$. However, \textsc{Cloudy} only includes cosmic ray heating of H\textsc{i}, hence the total heating rates in our model are higher than \textsc{Cloudy} in this region. The most important coolants in the fully shielded gas are molecular hydrogen and CO. 

\begin{figure*}
\centering
\mbox{
	\includegraphics[width=168mm]{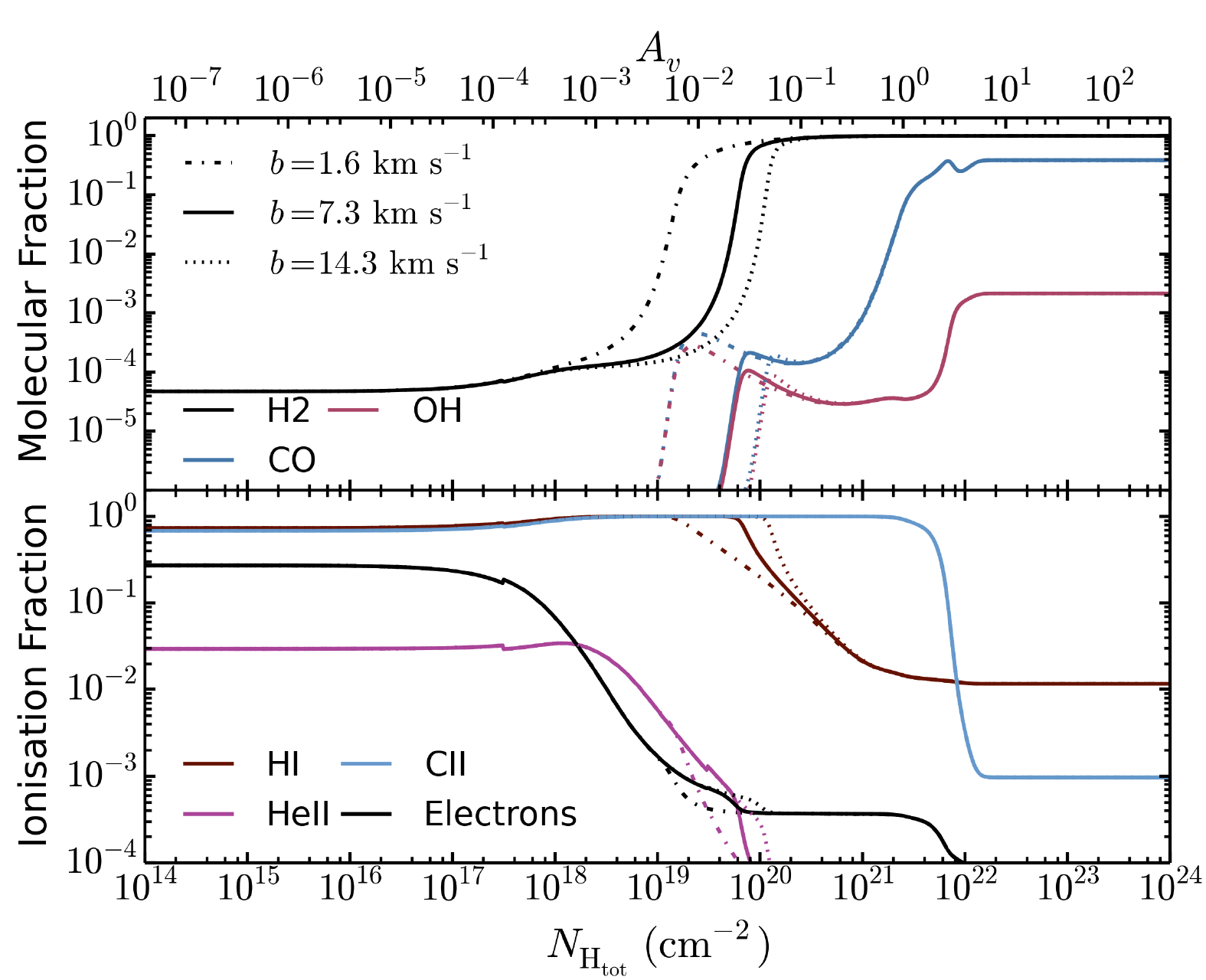}}
\caption{Comparison of the chemical properties of a one-dimensional plane-parallel slab of gas for three different Doppler broadening parameters, corresponding to different levels of turbulence in the gas, with $b = 1.6 \, \text{km s}^{-1}$ corresponding to pure thermal broadening. \textbf{Top panel:} equilibrium molecular fractions ($2 n_{\rm{H_{2}}} / n_{\rm{H_{tot}}}$, $n_{\rm{CO}} / n_{\rm{C_{tot}}}$ and $n_{\rm{OH}} / n_{\rm{O_{tot}}}$); \textbf{bottom panel:} equilibrium ionisation fractions. These were calculated at solar metallicity, a constant density $n_{\rm{H}} = 100 \, \text{cm}^{-3}$ and a constant temperature $T = 300$ K. The strongest turbulence shown here ($b = 14.3 \, \text{km s}^{-1}$) increases the column density at which the hydrogen starts to become molecular by more than an order of magnitude compared to pure thermal Doppler broadening, although its impact on the transition column density at which $x_{\rm{H_{2}}} = 0.5$ is less significant.}
\label{H2selfShieldFig}
\end{figure*}  

We have also compared our abundances and cooling and heating rates with \textsc{Cloudy} at densities of $1 \, \text{cm}^{-3}$ and $10^{4} \, \text{cm}^{-3}$, and a temperature of 100 K. These results can be found on our website. We generally find good agreement with \textsc{Cloudy}, although we sometimes find that the abundances of singly ionised metals with low ionisation energies, such as Si\textsc{ii} and Fe\textsc{ii}, are in poor agreement in fully shielded gas, once carbon becomes fully molecular. This occurs at the highest densities that we consider, although we begin to see such discrepancies in the Si\textsc{ii} abundance in figure~\ref{shieldedCombinedFig}. In this regime the ionisation of these species is typically dominated by charge transfer reactions between metal species, the rates of which tend to be uncertain. Moreover, a significant fraction of Si is found in SiO in the \textsc{Cloudy} models and, because our simplified molecular network does not include silicon molecules, we are unable to reproduce the correct silicon abundances in fully shielded gas. Since the abundances of other low ionisation energy ions such as Fe\textsc{ii} and Mg\textsc{ii} are dependent on Si\textsc{ii} due to charge transfer reactions in this regime, these species will also be uncertain in fully shielded gas. However, the predicted abundances of molecules such as H$_{2}$ and CO are in good agreement with \textsc{Cloudy} in this regime. 

\begin{figure*}
\centering
\mbox{
	\includegraphics[width=150mm]{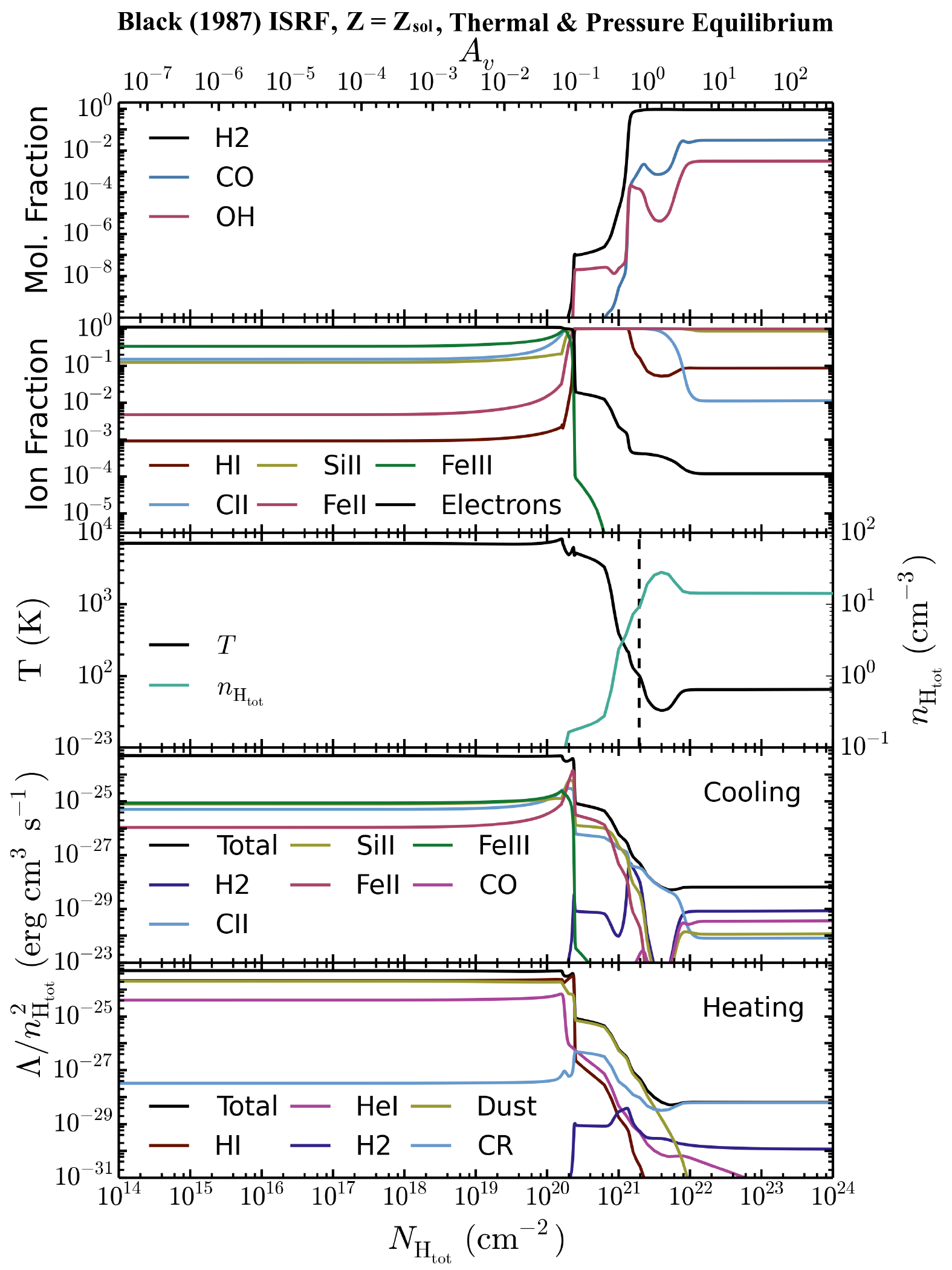}}
\caption{Chemistry and cooling properties of a one-dimensional plane-parallel slab of gas that is in thermal and pressure equilibrium with a pressure $P / k_{B} = 10^{3} \, \text{cm}^{-3} \, \text{K}$, plotted as a function of the total hydrogen column density into the gas cloud, assuming solar metallicity in the presence of the \citet{black87} interstellar radiation field. \textbf{Top panel:} equilibrium molecular fractions ($2 n_{\rm{H_{2}}} / n_{\rm{H_{tot}}}$, $n_{\rm{CO}} / n_{\rm{C_{tot}}}$ and $n_{\rm{OH}} / n_{\rm{O_{tot}}}$); \textbf{second panel:} equilibrium ionisation fractions; \textbf{third panel:} equilibrium gas temperature and density; \textbf{fourth panel:} equilibrium cooling rates; \textbf{bottom panel:} equilibrium heating rates. The vertical dashed line in the third panel indicates the column density at which the gas becomes Jeans unstable. To the right of this dashed line we would expect the gas cloud to become self-gravitating, and the density profile that we have imposed assuming pressure equilibrium will underestimate the typical densities corresponding to these column densities. At low column densities, the total cooling rate in the fourth panel is dominated by O\textsc{iii} and S\textsc{iii} (not shown here).}
\label{ShieldIsobaricFig}
\end{figure*}  

\subsection{Importance of turbulence for H$_{2}$ self-shielding}\label{H2selfSection}

In the previous section we found that the small H$_{2}$ abundance in the photodissociated region can be sufficient to begin attenuating the photodissociation rate of molecular hydrogen via self-shielding before dust extinction becomes significant. However, the presence of turbulence in the gas can suppress self-shielding due to Doppler broadening of the Lyman Werner lines. In this section we investigate the impact that turbulence can have on the H\textsc{i}-H$_{2}$ transition by repeating the above calculations for three different values of the turbulent Doppler broadening parameter: $b_{\rm{turb}} = 0 \, \text{km} \, \text{s}^{-1}$, $b_{\rm{turb}} = 7.1 \, \text{km} \, \text{s}^{-1}$ (our default value) and $b_{\rm{turb}} = 14.2 \, \text{km} \, \text{s}^{-1}$. The thermal Doppler broadening parameter at 300 K is $b_{\rm{therm}} = 1.6 \, \rm{km} \, \rm{s}^{-1}$, thus the total Doppler broadening parameters are $b = 1.6 \, \text{km} \, \text{s}^{-1}$, $7.3 \, \text{km} \, \text{s}^{-1}$ and $14.3 \, \text{km} \, \text{s}^{-1}$.

In figure~\ref{H2selfShieldFig} we show the molecular and ionisation fractions for these three Doppler broadening parameters at a constant density $n_{\rm{H_{tot}}} = 100 \, \text{cm}^{-3}$ and a constant temperature $T = 300$ K. We find that, while the ion fractions are hardly affected, increasing the turbulent Doppler broadening parameter from 0 to 14.2 km s$^{-1}$ increases the column density at which the hydrogen starts to become molecular by more than an order of magnitude, as it suppresses H$_{2}$ self-shielding. However, the transition also becomes sharper for higher values of $b$, so the change in the column density at which half of the hydrogen is molecular is less significant, increasing from $N_{\rm{H_{tot}}} \approx 3.3 \times 10^{19} \, \text{cm}^{-2}$ to $N_{\rm{H_{tot}}} \approx 1.4 \times 10^{20} \, \text{cm}^{-2}$. The transition to molecular hydrogen is still caused by H$_{2}$ self-shielding in these examples, even for the highest value of $b$ that we consider here.

\subsection{Atomic to molecular transition in thermal and pressure equilibrium}\label{isobarShielding}

The previous sections considered gas at a constant temperature of 300 K, but figure~\ref{shieldedCombinedFig} showed that the heating and cooling rates vary stongly with the column density into the cloud. Furthermore, we assumed a constant density throughout the cloud, which is unrealistic as gas at low column densities will be strongly photoheated and will thus typically have lower densities. It would therefore be more realistic to consider a cloud that is in thermal and pressure equilibrium. To achieve this, we first ran a series of models with different constant densities that were allowed to evolve to thermal equilibrium, from which we obtained the thermal equilibrium temperature on a two-dimensional grid of density and column density. Then, for each column density bin, we used this grid to determine the density that will give the assumed pressure. Finally, we imposed this isobaric density profile on the one-dimensional plane-parallel slab of gas and evolved it until it reached thermal and chemical equilibrium. This scenario is more typical of the two-phase ISM, in which gas is photoheated to higher temperatures, with lower densities, at low column densities, and then cools to a much colder and denser phase once the ionising photons have been attenuated and molecular cooling becomes important. 

In figure~\ref{ShieldIsobaricFig} we show the results for a cloud with a constant pressure $P / k_{\rm{B}} = 10^{3} \, \rm{cm}^{-3} \, \rm{K}$, assuming solar metallicity. The equilibrium molecular and ionisation fractions of some species are shown in the top two panels of figure~\ref{ShieldIsobaricFig} as a function of the total column density $N_{\rm{H_{tot}}}$, the equilibrium gas temperature and density are shown in the middle panel, and the equilibrium cooling and heating rates are shown in the bottom two panels. 

We see that gas is more strongly ionised in the photodissociated region compared to the constant temperature and density run in figure~\ref{shieldedCombinedFig}. This is due to both the higher temperature and the lower density. Furthermore, the molecular hydrogen fraction is much lower in this region, and the H$_{2}$ self-shielding is thus weaker. Hence, the H\textsc{i}-H$_{2}$ transition occurs at a higher column density compared to figure~\ref{shieldedCombinedFig} ($x_{\rm{H_{2}}} = 0.5$ at $N_{\rm{H_{tot}}} \approx 1.5 \times 10^{21} \, \rm{cm}^{-2}$ in figure~\ref{ShieldIsobaricFig}, compared to $N_{\rm{H_{tot}}} \approx 8.1 \times 10^{19} \, \rm{cm}^{-2}$ in figure~\ref{shieldedCombinedFig}). The H$_{2}$ fraction first rises at $N_{\rm{H_{tot}}} \sim 2 \times 10^{20} \, \text{cm}^{-2}$ due to a corresponding rise in the H\textsc{i} abundance, which is required to form H$_{2}$. The H\textsc{i} abundance increases because it is able to shield itself against ionising radiation above 1 Ryd (the H\textsc{i} column density here is $\sim 10^{17} \, \text{cm}^{-2}$). However, the H$_{2}$ fraction still remains low after this initial increase ($x_{\rm{H_{2}}} \approx 10^{-7}$). The H\textsc{i}-H$_{2}$ transition at $N_{\rm{H_{tot}}} \approx 1.5 \times 10^{21} \, \rm{cm}^{-2}$ is initially triggered by the increasing density, which enhances the formation rate of H$_{2}$ on dust grains with respect to the photodissociation rate. However, H$_{2}$ self-shielding then becomes significant and is responsible for the final transition to a fully molecular gas. When we repeat this model without H$_{2}$ self-shielding, we find that the H\textsc{i}-H$_{2}$ transition occurs at a factor $\sim \! 5$ higher column density, with $x_{\rm{H_{2}}} = 0.5$ at $N_{\rm{H_{tot}}} \approx 7.2 \times 10^{21} \, \rm{cm}^{-2}$.

In figure~\ref{ShieldIsobaricFig}, CO becomes fully shielded from dissociating radiation at $N_{\rm{H_{tot}}} \ga 10^{22} \, \rm{cm}^{-2}$. Like in the constant density run in figure~\ref{shieldedCombinedFig}, this transition is determined by dust shielding. However, the fraction of carbon in CO in the fully shielded region is $\la 5$ per cent in figure~\ref{ShieldIsobaricFig}, compared to $\sim 40$ per cent in figure~\ref{shieldedCombinedFig}, with most of the remaining carbon in C\textsc{i} (not shown in the figures). This lower abundance of CO in the isobaric run is due to the lower density, which is a factor of $\sim 10$ lower than the constant density run in the fully shielded region. 

Observations of diffuse clouds at low column densities find abundances of CH$^{+}$ that are several orders of magnitude higher than can be explained using standard chemical models \citep[e.g.][]{federman96,sheffer08,visser09}. This also leads to predicted CO abundances that are lower than observed in such regions, as CH$^{+}$ is an important formation channel for CO. Various non-thermal production mechanisms for CH$^{+}$ have been proposed to alleviate this discrepancy. For example, \citet{federman96} suggest that Alfv\'{e}n waves that enter the cloud can produce non-thermal motions between ions and neutral species, thus increasing the rates of ion-neutral reactions. 

To see what effect such suprathermal chemistry has on our chemical network, we repeated the models in figures~\ref{shieldedCombinedFig} and \ref{ShieldIsobaricFig} with the kinetic temperature of all ion-neutral reactions replaced by an effective temperature that is enhanced by Alfv\'{e}n waves, as given by the prescription of \citet{federman96} with an Alfv\'{e}n speed of $3.3 \, \rm{km} \, \rm{s}^{-1}$. Following \citet{sheffer08} and \citet{visser09}, we only include this enhancement of the ion-neutral reactions at low column densities, $N_{\rm{H_{2}}} < 4 \times 10^{20} \, \rm{cm}^{-2}$. In the model with a constant density $n_{\rm{H_{tot}}} = 100 \, \rm{cm}^{-3}$ and constant temperature $T = 300 \, \rm{K}$, we find that the CH$^{+}$ abundance is increased by up to four orders of magnitude at intermediate column densities $5 \times 10^{19} \, \rm{cm}^{-2} \la N_{\rm{H_{tot}}} \la 10^{21} \, \rm{cm}^{-2}$, which is consistent with previous studies \citep[e.g.][]{federman96}. However, the CO abundance is only enhanced by up to a factor of 6 in this same region, which is less than the enhancements in CO abundance seen in \citet[e.g.][]{sheffer08}, who find that the CO abundance increases by a factor $\sim 100$ for an Alfv\'{e}n speed of $3.3 \, \rm{km} \, \rm{s}^{-1}$. We see similar enhancements in the model with a constant pressure $P / k_{\rm{B}} = 10^{3} \, \rm{cm}^{-3} \, \rm{K}$, but only at column densities $10^{21} \, \rm{cm}^{-2} \la N_{\rm{H_{tot}}} \la 3 \times 10^{21} \, \rm{cm}^{-2}$, as the temperature at lower column densities is much higher than in the constant density model ($T > 1000 \, \rm{K}$).

At low column densities the cooling rate in figure~\ref{ShieldIsobaricFig} is determined by several ionised species including Si\textsc{ii}, Fe\textsc{ii}, Fe\textsc{iii} and C\textsc{ii}, along with N\textsc{iii}, O\textsc{iii}, Ne\textsc{iii} and S\textsc{iii} (not shown in figure~\ref{ShieldIsobaricFig}). Heating at low column densities is primarily from dust heating and photoionisation of H\textsc{i} and He\textsc{i}. At $N_{\rm{H_{tot}}} \sim 2 \times 10^{20} \, \text{cm}^{-2}$ the cooling rates from Si\textsc{ii} and Fe\textsc{ii} peak as Si\textsc{iii} and Fe\textsc{iii} recombine, creating a small dip in the temperature profile. The photoheating rate then drops sharply as the ionising radiation above 1 Ryd becomes shielded by H\textsc{i}, and the total heating rate becomes dominated by the dust photoelectric effect. The thermal equilibrium temperature reaches a minimum of 30 K at $N_{\rm{H_{tot}}} \sim 4 \times 10^{21} \, \text{cm}^{-2}$ as the dust photoelectric heating becomes suppressed by dust shielding, leaving heating primarily from cosmic ray ionisation of H$_{2}$. However, after this point the carbon forms CO and the cooling becomes dominated by molecules (CO and H$_{2}$). These species are less efficient at cooling than C\textsc{ii}, so the thermal equilibrium temperature increases to 60 K. Hence, the temperature in the fully molecular region is higher than the minimum at $N_{\rm{H_{tot}}} \sim 4 \times 10^{21} \, \text{cm}^{-2}$. 

We note that the assumption of pressure equilibrium in these examples will be unrealistic at high column densities, where the gas becomes self-gravitating. The vertical dashed line in the middle panel of figure~\ref{ShieldIsobaricFig} indicates the column density at which the size of the system, $N_{\rm{H_{tot}}} / n_{\rm{H_{tot}}}$, becomes larger than the local Jeans length, at $N_{\rm{H_{tot}}} \approx 2 \times 10^{21} \, \text{cm}^{-2}$. Thus, we would expect the fully molecular region in these examples to have a higher density than we have used here. 

We also considered a gas cloud with a pressure that is a factor 100 larger than is shown in figure~\ref{ShieldIsobaricFig}. This results in higher densities in the fully shielded region that are more typical of molecular clouds ($n_{\rm{H_{tot}}} \sim 10^{3} \, \rm{cm}^{-3}$), although the densities in the photodissociated region are then higher than we would expect for the warm ISM. The cumulative H$_{2}$ fractions for the two pressures are shown in figure~\ref{H2modelsFig} (see section~\ref{H2modelsSection}). The top panels are at the lower pressure ($P / k_{\rm{B}} = 10^{3} \, \text{cm}^{-3} \, \text{K}$), and the bottom panels are at the higher pressure ($P / k_{\rm{B}} = 10^{5} \, \text{cm}^{-3} \, \text{K}$). In the left panels we use a metallicity $0.1 Z_{\odot}$, and in the right panels we use solar metallicity. We find that in all of these examples H$_{2}$ self-shielding is again important for the final transition to fully molecular hydrogen, reducing the column density of this transition by up to two orders of magnitude compared to the same model without H$_{2}$ self-shielding. 

We thus find that the effect of H$_{2}$ self-shielding can be weaker in a cloud that is in thermal and pressure equilibrium due to the lower densities at low column densities compared to a model with a constant density and temperature. However, H$_{2}$ self-shielding still determines the final transition to fully molecular hydrogen. Furthermore, H$_{2}$ self-shielding will be even more important if the pressure (and hence densities) are higher. This trend with density agrees with previous models of photodissociation regions and molecular clouds \citep[e.g.][]{black87b,krumholz09}. 

Figure~\ref{H2modelsFig} also demonstrates the dependence of the transition column density on pressure and metallicity. The H\textsc{i}-H$_{2}$ transition occurs at a lower column density for higher pressure and higher metallicity. Both of these trends are driven by H$_{2}$ self-shielding because the H$_{2}$ fraction in the dissociated region is higher at higher pressure (due to the increased density) and at higher metallicity (due to the increased formation of H$_{2}$ on dust grains, and also due to the increased cooling from metals, which results in a lower temperature and a higher density at a given pressure). These trends with pressure (or density) and metallicity have also been seen in previous studies of the H$_{2}$ transition in photodissociation regions \citep[e.g.][]{wolfire10}. 

We also see the time evolution of the molecular hydrogen fraction, starting from neutral, atomic gas, in figure~\ref{H2modelsFig}. In the low pressure run at a metallicity $0.1 Z_{\odot}$ \textit{(top left panel)} the H$_{2}$ abundance in the fully shielded region takes $\sim 1$ Gyr to reach equilibrium. This time-scale is shorter in the high pressure runs and at higher metallicity. At the high pressure and solar metallicity \textit{(bottom right panel)}, the H$_{2}$ abundance already reaches equilibrium after $\sim 1$ Myr.

\section{Comparison with published approximations for H$_{2}$ formation}\label{H2modelsSection}

The connection between gas surface density and star formation rate density is well established observationally, both averaged over galactic scales \citep[e.g.][]{kennicutt89,kennicutt98} and also in observations that are spatially resolved on $\sim$kpc scales \citep{wong02,heyer04,schuster07}, although at smaller scales, comparable to giant molecular clouds ($\la 100$ pc), this relation breaks down \citep[e.g.][]{onodera10}. It has emerged more recently that star formation correlates more strongly with the molecular than with the atomic or total gas content \citep[e.g.][]{kennicutt07,leroy08,bigiel08,bigiel10}, although the more fundamental correlation may in fact be with the cold gas content \citep{schaye04,krumholz11a,glover12}. 

This important observational link between molecular gas and star formation has motivated several new models and prescriptions for following the H$_{2}$ fraction of gas in hydrodynamic simulations. These studies aim to use a more physical prescription for star formation and to investigate its consequence for galactic environments that are not covered by current observations, such as very low metallicity environments at high redshifts. Some of these prescriptions utilise very simple chemical models that include the formation of molecular hydrogen on dust grains and its photodissociation by Lyman Werner radiation to follow the non-equilibrium H$_{2}$ fraction \citep[e.g.][]{gnedin09,christensen12}. Others use approximate analytic models to predict the H$_{2}$ content of a cloud from its physical parameters such as the dust optical depth and incident photodissociating radiation field \citep[e.g.][]{krumholz08,krumholz09,mckee10}, which can then be applied to gas particles/cells in hydrodynamic simulations \citep[e.g.][]{krumholz11b,halle13}.

In this section we compare the molecular hydrogen fractions predicted by some of these models with our chemical network. We begin by introducing the molecular hydrogen models from the literature that we will compare our chemical network to, and then we present our results. 

\subsection{Gnedin09 model}\label{gnedin09Section}

\citet{gnedin09} present a simple prescription to follow the non-equilibrium evolution of the molecular hydrogen fraction in hydrodynamic simulations, which has been implemented in cosmological Adaptive Mesh Refinement (AMR) simulations \citep{gnedin10}, and was also adapted by \citet{christensen12} for cosmological Smoothed Particle Hydrodynamics (SPH) simulations. This model includes the formation of H$_{2}$ on dust grains, photodissociation by Lyman Werner radiation, shielding by both dust and H$_{2}$, and a small number of gas phase reactions that become important in the dust-free regime (we use the five gas phase reactions suggested by \citealt{christensen12}; see below). The evolution of the neutral and molecular hydrogen fractions, $x_{\rm{HI}}$ and $x_{\rm{H_{2}}}$, can then be described by the following rate equations:

\begin{align}
\dot{x}_{\rm{HI}} &= R(T) n_{e} x_{\rm{HII}} - S_{\rm{d}} x_{\rm{HI}} \Gamma_{\rm{HI}} - C_{\rm{HI}} x_{\rm{HI}} n_{\rm{e}} - 2 \dot{x}_{\rm{H_{2}}} , \\
\dot{x}_{\rm{H_{2}}} &= R_{\rm{d}} n_{\rm{H_{tot}}} x_{\rm{HI}} (x_{\rm{HI}} + 2 x_{\rm{H_{2}}}) - S_{\rm{d}} S_{\rm{self}}^{\rm{H_{2}}} \Gamma_{\rm{H_{2}}}^{\rm{LW}} + \dot{x}_{\rm{H_{2}}}^{\rm{gp}}. 
\end{align}
Following \citet{christensen12}, we have also included the collisional ionisation of H\textsc{i}, $C_{\rm{HI}}$, in the above equations, even though it was omitted by \citet{gnedin09}. $R(T)$ is the recombination rate of H\textsc{ii}, $\Gamma_{\rm{HI}}$ is the photoionisation rate of H\textsc{i} and $\Gamma_{\rm{H_{2}}}^{\rm{LW}}$ is the photodissociation rate of H$_{2}$ by Lyman Werner radiation. For the gas phase reactions, $\dot{x}_{\rm{H_{2}}}^{\rm{gp}}$, we include the five reactions suggested by \citet{christensen12}: the formation of H$_{2}$ via H$^{-}$ and its collisional dissociation via H$_{2}$, H\textsc{i}, H\textsc{ii} and $e^{-}$, with the abundance of H$^{-}$ assumed to be in chemical equilibrium, as given by equation (27) of \citet{abel97}. 

\citet{gnedin09} use the H$_{2}$ self-shielding factor $S_{\rm{self}}^{\rm{H_{2}}}$ from \citet{draine96}, albeit with different parameters. They find that using $\omega_{\rm{H_{2}}} = 0.2$ and a constant Doppler broadening parameter of $b = 1 \, \rm{km} \, \rm{s}^{-1}$ in equation~\ref{H2self_eqn}, along with the original value of $\alpha = 2$, produces better agreement between their model and observations. This choice of parameters results in weaker H$_{2}$ self-shielding compared to our temperature-dependent self-shielding function (equations~\ref{H2self_mod_eqn} to \ref{NcritEqn}) for the same value of $b$. However, in our fiducial model we include a turbulent Doppler broadening parameter of $b_{\rm{turb}} = 7.1 \, \rm{km} \, \rm{s}^{-1}$, which makes the self-shielding weaker in our model than in the Gnedin09 model at intermediate H$_{2}$ column densities ($10^{14} \, \rm{cm}^{-2} \la N_{\rm{H_{2}}} \la 10^{16} \, \rm{cm}^{-2}$).

The dust shielding factor $S_{\rm{d}}$ is given in equation~\ref{dust_shielding}, except that \citet{gnedin09} use an effective dust area per hydrogen atom of $\sigma_{\rm{d, eff}} = \gamma_{\rm{d}}^{\rm{H_{2}}} A_{v} / N_{\rm{H_{tot}}} = 4 \times 10 ^{-21} \, \rm{cm}^{2}$. This is a factor of 2.7 larger than the value that we use. Also, unlike us, they use the same dust shielding factor for both H$_{2}$ and H\textsc{i}.

For the rate of formation of H$_{2}$ on dust grains ($R_{\rm{d}}$), \citet{gnedin09} use a rate that was derived from observations by \citet{wolfire08}, scaled linearly with the metallicity and multiplied by a clumping factor $C_{\rho}$ that accounts for the fact that there may be structure within the gas below the resolution limit, and that molecular hydrogen will preferentially form in the higher density regions:

\begin{equation}\label{gnedinRateEqn}
R_{\rm{d}} = 3.5 \times 10^{-17} Z / Z_{\odot} C_{\rho} \, \text{cm}^{3} s^{-1}.
\end{equation}
\citet{gnedin09} use a clumping factor $C_{\rho} = 30$, but in our comparisons we consider spatially resolved, one-dimensional simulations of an illuminated slab of gas. Therefore, for these comparisons we shall take $C_{\rho} = 1$. For gas temperatures $10 \, \text{K} < T < 10^{3} \, \text{K}$, the rate in equation~\ref{gnedinRateEqn} is within a factor $\sim 2$ of the value used in our model, taken from \citet{cazaux02} with a dust temperature of 10 K. However, above $10^{3}$ K the H$_{2}$ formation rate on dust grains in our model decreases. This temperature dependence is not included in the Gnedin09 model. 

Finally, the abundances of electrons and H\textsc{ii} can be calculated from constraint equations. 

\citet{gnedin11} expand the Gnedin09 model to include the Helium chemistry. There are also other examples in the literature of methods that follow the non-equilibrium evolution of H$_{2}$ using simplified chemical models. For example, \citet{bergin04} present a model for the evolution of molecular hydrogen that includes the formation of H$_{2}$ on dust grains and its dissociation by Lyman Werner radiation and cosmic rays, although they do not include the gas phase reactions (see their equation A1). The model of \citet{bergin04} has been used to investigate the formation of molecular clouds in galactic discs \citep[e.g.][]{dobbs08,khoperskov13}. We only consider the Gnedin09 chemical model in this section.

\subsection{KMT model}\label{KMT_model}
\citet{krumholz08,krumholz09} and \citet{mckee10} develop a simple analytic model that considers a spherical gas cloud that is immersed in an isotropic radiation field. By solving approximately the radiative transfer equation with shielding of the Lyman Werner radiation by dust and H$_{2}$ to obtain the radial dependence of the radiation field, and then balancing the photodissociation of molecular hydrogen against its formation on dust grains, they derive simple analytic estimates for the size of the fully molecular region of the cloud, and hence for its molecular fraction in chemical equilibrium. 

We use equation (93) of \citet{mckee10} to calculate the mean equilibrium H$_{2}$ fraction, $f_{\rm{H_{2}}}$, of a gas cloud:

\begin{equation}
f_{\rm{H_{2}}} = 1 - \left( \frac{3}{4} \right) \frac{s}{1 + 0.25 s},
\end{equation}
where $s$ is given by their equation (91):

\begin{equation}
s = \frac{\ln(1 + 0.6 \chi + 0.01 \chi^{2})}{0.6 \tau_{c}},
\end{equation}
where $\chi$ and $\tau_{c}$ are dimensionless parameters of their model, which represent a measure of the incident Lyman Werner radiation field and the dust optical depth to the centre of the cloud respectively. They are given by their equations (9) and (86):

\begin{equation}
\chi = 71 \left( \frac{\sigma_{\rm{d}, -21}}{R_{\rm{d}, -16.5}} \right) \frac{G^{'}_{0}}{(n_{\rm{H}} / \rm{cm}^{-3})},
\end{equation}

\begin{equation}
\tau_{c} = \frac{3}{4} \left( \frac{\Sigma \sigma_{\rm{d}}}{\mu_{\rm{H}}} \right) = \sigma_{\rm{d}} N_{\rm{H_{tot}}} ,
\end{equation}
where $\sigma_{\rm{d}, -21} = \sigma_{\rm{d}} / 10^{-21} \, \rm{cm}^{2}$, $\sigma_{\rm{d}}$ is the cross-sectional area of dust grains available to absorb photons (\citealt{krumholz09} use $\sigma_{\rm{d}} = 10^{-21} Z / Z_{\odot} \, \rm{cm}^{2}$), and $R_{\rm{d}, -16.5} = R_{\rm{d}} / 10^{-16.5} \rm{cm}^{3} \, \rm{s}^{-1}$, where $R_{\rm{d}}$ is the rate coefficient for H$_{2}$ formation on dust grains. For the latter, \citet{mckee10} use an observationally determined value from \citet{draine96}, multiplied by the metallicity: $R_{\rm{d}, -16.5} = Z / Z_{\odot}$. \citet{krumholz11b} also boost $R_{\rm{d}, -16.5}$ by the clumping factor $C_{\rho}$, whereas \citet{krumholz13} multiply $\tau_{c}$ by the clumping factor, but we take $C_{\rho} = 1$. $G^{'}_{0}$ is the number density of dissociating photons in the Lyman Werner band, normalised to the value from \citet{draine78} for the Milky Way. For the \citet{black87} interstellar radiation field that we consider here, $G^{'}_{0} = 0.803$. $n_{\rm{H}}$ is the total hydrogen number density, $\mu_{\rm{H}}$ is the mean mass per hydrogen nucleus, $\Sigma$ is the average surface density of the spherical cloud and $N_{\rm{H_{tot}}}$ is the total hydrogen column density from the edge of the cloud to its centre. 

\citet{krumholz09} and \citet{krumholz13} demonstrate that the molecular gas content predicted by this model is in agreement with various extragalactic observations and with observations of molecular clouds within the Milky Way, for particular values of the clumping factor $C_{\rho}$ and assumptions about the radiation field and/or density. For example, \citet{krumholz09} assume that the ratio of the radiation field to the density, $G^{'}_{0} / n_{\rm{H}}$, is set by the minimum density of the cold neutral medium required for the ISM to be in two-phase equilibrium. As shown in their equation 7, the ratio $G^{'}_{0} / n_{\rm{H}}$ then depends only on metallicity under this assumption. \citet{krumholz13} extends the approach of \citet{krumholz09} by assuming that, as $G^{'}_{0} \to 0$, the density $n_{\rm{H}}$ reaches a minimum floor, which is required to ensure that the thermal pressure of the ISM is able to maintain hydrostatic equilibrium in the disc. This becomes important in the molecule-poor regime. In contrast, the runs that we present below use a constant incident radiation field (the \citealt{black87} interstellar radiation field) and assume a density profile such that the gas pressure is constant throughout the cloud. 

\citet{krumholz11b} implement this model in cosmological AMR simulations to estimate the equilibrium H$_{2}$ fraction of each gas cell. \citet{halle13} also implement the KMT model in SPH simulations of isolated disc galaxies to investigate the role that the cold molecular phase of the ISM plays in star formation and as a gas reservoir in the outer disc. 

\subsection{Results}

\begin{figure*}
\centering
\mbox{
	\includegraphics[width=168mm]{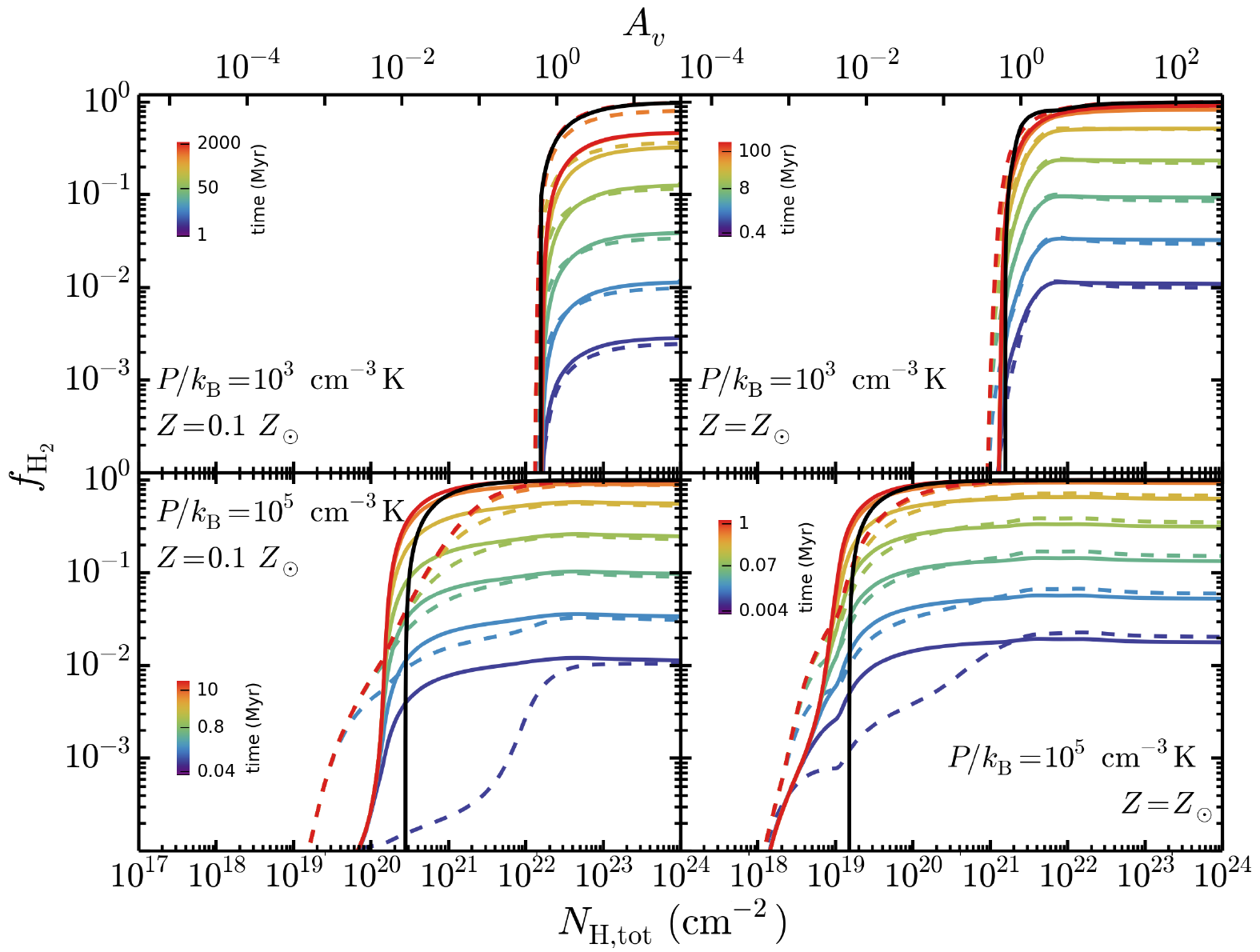}}
\caption{Comparison of the cumulative molecular hydrogen fraction, plotted as a function of total hydrogen column density, as predicted by our chemical model \textit{(solid coloured curves)}, the Gnedin09 model \textit{(dashed coloured curves)} and the equilibrium KMT model \textit{(solid black curve)}. The coloured lines show the non-equilibrium evolution measured at logarithmic time intervals, as indicated by the colour bars, until the simulation reaches chemical equilibrium (red curves). Each run was calculated with a density and temperature profile in pressure and thermal equilibrium, with a pressure $P / k_{\rm{B}} = 10^{3} \, \rm{cm}^{-3} \, \rm{K}$ \textit{(top row)} and $P / k_{\rm{B}} = 10^{5} \, \rm{cm}^{-3} \, \rm{K}$ \textit{(bottom row)}. We used a metallicity $0.1 Z_{\odot}$ \textit{(left column)} and $Z_{\odot}$ \textit{(right column)}. At low pressure the agreement between the different models is good, except that in the low metallicity run (top left panel), cosmic ray dissociation of H$_{2}$, which was not included in the KMT and Gnedin09 models, reduces the equilibrium H$_{2}$ fraction by a factor of two in the fully shielded region. At high pressure the differences between the models are substantial.}
\label{H2modelsFig}
\end{figure*}  

In Figure~\ref{H2modelsFig} we compare the molecular abundances from our model \textit{(coloured solid curves)} with the simpler Gnedin09 \textit{(coloured dashed curves)} and KMT \textit{(black solid curves)} models described above. These were calculated using temperature and density profiles that are in thermal and pressure equilibrium. We use two different pressures, $P / k_{\rm{B}} = 10^{3} \, \rm{cm}^{-3} \, \rm{K}$ \textit{(top row)} and $P / k_{\rm{B}} = 10^{5} \, \rm{cm}^{-3} \, \rm{K}$ \textit{(bottom row)}, and two metallicities, $Z = 0.1 Z_{\odot}$ \textit{(left column)} and $Z_{\odot}$ \textit{(right column)}. The colour encodes the time evolution in our model and the Gnedin09 model, starting from fully neutral, atomic gas (KMT is an equilibrium model). In figures~\ref{shieldedCombinedFig}-\ref{ShieldIsobaricFig} we showed the abundances in each gas cell, but figure~\ref{H2modelsFig} shows the cumulative molecular fraction, i.e. the fraction of all hydrogen atoms that is in H$_{2}$ up to a given column density. This allows us to compare our results to the KMT model, in which the specified column density is measured to the centre of a spherical cloud, and then the H$_{2}$ fraction, $f_{\rm{H_{2}}}$, is the fraction of gas in the entire cloud that is molecular, rather than the fraction of molecular gas at that column density. However, this is still not entirely equivalent to our model as we assume a plane-parallel geometry, whereas the KMT model assumes a spherical geometry. 

In the low pressure runs, the final molecular fractions predicted by our model generally agree well with both the Gnedin09 and the KMT models. The time evolution of the H$_{2}$ fraction in our model and the Gnedin09 model are also in good agreement. The largest discrepancy in the equilibrium abundance is in the fully shielded region at low metallicity (top left panel of figure~\ref{H2modelsFig}), where the equilibrium H$_{2}$ fraction in our model is a factor of two lower than that predicted by the Gnedin09 and KMT models. This is due to cosmic ray dissociation of H$_{2}$, which lowers the H$_{2}$ abundance but is not included in the Gnedin09 or KMT models. 

In the high pressure runs the H\textsc{i}-H$_{2}$ transition occurs at a slightly lower column density than in the KMT model. For example, at solar metallicity (bottom right panel of figure~\ref{H2modelsFig}), we predict that $f_{\rm{H_{2}}} = 0.5$ at $N_{\rm{H_{tot}}} = 2.2 \times 10^{19} \, \rm{cm}^{-2}$, compared to $N_{\rm{H_{tot}}} = 3.0 \times 10^{19} \, \rm{cm}^{-2}$ for KMT. In the Gnedin09 model, the H\textsc{i}-H$_{2}$ transition is noticeably flatter at high pressure than in our model or in the KMT model. The H$_{2}$ fraction starts to increase at a lower column density in the Gnedin09 model than in the other two models, but the column density at which $f_{\rm{H_{2}}} = 0.5$ is higher, e.g. $N_{\rm{H_{tot}}} = 5.6 \times 10^{19} \, \rm{cm}^{-2}$ at solar metallicity. This difference is due to the different H$_{2}$ self-shielding function that is used in the Gnedin09 model.

\citet{krumholz11b} compare the KMT and Gnedin09 models in a hydrodynamic simulation of a Milky Way progenitor galaxy. For the Gnedin09 model, they include the changes described in \citet{gnedin11}, which adds Helium chemistry to the chemical network, and uses the simpler power-law H$_{2}$ self-shielding function given in equation 36 of \citet{draine96}. They find that the molecular fractions predicted by the two models are in excellent agreement for metallicities $Z > 10^{-2} \, Z_{\odot}$, with discrepancies at lower metallicities likely due to time-dependent effects. This is consistent with what we find in the top row of figure~\ref{H2modelsFig} for the low pressure runs. In contrast, the high pressure runs in the bottom row of figure~\ref{H2modelsFig} show poor agreement between the KMT and Gnedin09 models. However, this high pressure was chosen to reproduce densities typical of molecular clouds ($n_{\rm{H_{tot}}} \sim 10^{3} \, \rm{cm}^{-2}$) in the fully shielded region, and it produces unusually high densities in the low column density region ($n_{\rm{H_{tot}}} \sim 10 \, \rm{cm}^{-2}$). Such regions of high density and low column density were not probed by the simulation of \citet{krumholz11b}, and are likely to be rare in realistic galactic environments. Therefore, the discrepancies that we see in figure~\ref{H2modelsFig} do not contradict the results of \citet{krumholz11b}.

In all three models the H\textsc{i}-H$_{2}$ transition occurs at a lower column density at higher metallicity and at higher pressure. As described in section~\ref{isobarShielding}, these trends are driven by H$_{2}$ self-shielding, because the H$_{2}$ fraction in the photodissociated region is higher at high metallicity and at high pressure. These trends have also been seen in previous models of photodissociation regions \citep[e.g.][]{wolfire10}. From the colour bars in each panel of figure~\ref{H2modelsFig}, we also see that the molecular fractions reach chemical equilibrium faster at higher metallicity and higher pressure.

To confirm the impact that H$_{2}$ self-shielding has on the H\textsc{i}-H$_{2}$ transition in our model, we repeated the above calculations with H$_{2}$ self-shielding switched off. We find that the total hydrogen column density of the H\textsc{i}-H$_{2}$ transition, at which $f_{\rm{H_{2}}} = 0.5$, increases significantly when H$_{2}$ self-shielding is omitted, for example by a factor of $\sim 5$ and $\sim 300$ in the low and high pressure runs respectively at solar metallicity. This confirms that the H\textsc{i}-H$_{2}$ transition is determined by H$_{2}$ self shielding in all of the examples shown in figure~\ref{H2modelsFig}. The importance of H$_{2}$ self-shielding for the H\textsc{i}-H$_{2}$ in photodissociation region models has previously been studied by e.g. \citet{black87b,draine96,lee96}. 

\section{Conclusions}\label{conclusions}

We have extended the thermo-chemical model from paper I to account for gas that becomes shielded from the incident UV radiation field. We attenuate the photoionisation, photodissociation and photoheating rates by dust and by the gas itself, including absorption by H\textsc{i}, H$_{2}$, He\textsc{i}, He\textsc{ii} and CO where appropriate. For the self-shielding of H$_{2}$, we use a new temperature-dependent analytic approximation that we fit to the suppression of the H$_{2}$ photodissociation rate predicted by \textsc{Cloudy} as a function of H$_{2}$ column density (see appendix~\ref{H2self_comparison_section}). Using this model, we investigated the impact that shielding of both the photoionising and the photodissociating radiation has on the chemistry and the cooling properties of the gas. 

We have performed a series of one-dimensional calculations of a plane-parallel slab of gas illuminated by the \citet{black87} interstellar radiation field at constant density. Comparing equilibrium abundances and cooling and heating rates as a function of column density with \textsc{Cloudy}, we generally find good agreement. At $n_{\rm{H_{tot}}} = 100 \, \text{cm}^{-3}$, solar metallicity and a constant temperature $T = 300$ K, we find that the H\textsc{i}-H$_{2}$ transition occurs at a somewhat lower column density in our model than in \textsc{Cloudy}'s big H2 model, with a molecular hydrogen fraction $x_{\rm{H_{2}}} = 0.5$ at $N_{\rm{H_{tot}}} \approx 8.1 \times 10^{19} \, \rm{cm}^{-2}$ in our model, compared to $N_{\rm{H_{tot}}} \approx 2.8 \times 10^{20} \, \rm{cm}^{-2}$ in \textsc{Cloudy} (see figure~\ref{shieldedCombinedFig}). However, \textsc{Cloudy}'s small H2 model predicts an H\textsc{i}-H$_{2}$ transition column density that is closer to our value, with $x_{\rm{H_{2}}} = 0.5$ at $N_{\rm{H_{tot}}} \approx 1.1 \times 10^{20} \, \rm{cm}^{-2}$. 

In the examples shown here, the H\textsc{i}-H$_{2}$ transition is determined by H$_{2}$ self-shielding in our model, as the residual molecular hydrogen fraction in the photodissociated region at low column densities is sufficient for self-shielding to become important before dust shielding. As the photodissociation rate is slightly lower in our model than in the big H2 model of \textsc{Cloudy}, the H$_{2}$ fraction at low column densities is slightly higher in our model. This explains why the transition column density is somewhat lower in our model compared to the \textsc{Cloudy} big H2 model. The importance of H$_{2}$ self-shielding also means that the molecular hydrogen transition is sensitive to turbulence, which can suppress self-shielding. The transition for carbon to form CO is primarily triggered by dust shielding and thus occurs at a higher column density, $N_{\rm{H_{tot}}} \sim 10^{22} \, \text{cm}^{-2}$. 

We also consider gas clouds with temperature and density profiles that are in thermal and pressure equilibrium, which is more realistic for a two-phase ISM than a gas cloud with constant temperature and density throughout. The effect of H$_{2}$ self-shielding is weaker in these examples due to the lower densities in the photodissociated region (see figure~\ref{ShieldIsobaricFig}). However, the H\textsc{i}-H$_{2}$ transition is still determined by self-shielding, which becomes more important in runs with higher pressure (and hence higher densities). This trend with density is consistent with previous studies that have looked at the importance of H$_{2}$ self-shielding for the H\textsc{i}-H$_{2}$ transition\citep[e.g.][]{black87b,draine96,lee96,krumholz09}. 

The H\textsc{i}-H$_{2}$ transition occurs at a lower total column density for higher density (or equivalently, for higher pressure if the cloud is in pressure equilibrium) and for higher metallicity (see figure~\ref{H2modelsFig}), in agreement with previous models of photodissociation regions \citep[e.g.][]{wolfire10}. These trends are due to the H$_{2}$ self-shielding, because an increase in the density and/or metallicity will increase the H$_{2}$ fraction in the photodissociated region, and hence decrease the total column density at which the H$_{2}$ becomes self-shielded. The time evolution of the H$_{2}$ fraction is also dependent on the density (or pressure) and the metallicity. For a gas cloud with pressure $P / k_{B} = 10^{3} \, \text{cm}^{-3} \, \text{K}$ and metallicity $0.1 Z_{\odot}$, the molecular hydrogen abundance in the fully shielded region only reaches equilibrium (starting from neutral, atomic initial conditions) after $\sim 1$ Gyr. This time-scale decreases as the pressure and/or the metallicity increase. 

We compare the dominant cooling and heating processes in our low pressure example ($P / k_{B} = 10^{3} \, \text{cm}^{-3} \, \text{K}$) at solar metallicity, and we find that they form three distinct regions (see figure~\ref{ShieldIsobaricFig}). At low column densities, where the dissociating and ionising radiation flux is still high, cooling is primarily from ionised metals such as Si\textsc{ii}, Fe\textsc{ii}, Fe\textsc{iii} and C\textsc{ii}, which are balanced by photoheating, primarily from H\textsc{i}. At column densities above $N_{\rm{H_{tot}}} \sim 2 \times 10^{20} \, \text{cm}^{-2}$ the hydrogen-ionising radiation above 1 Ryd becomes significantly attenuated by neutral hydrogen. This reduces the photoheating rates, making photoelectric dust heating the dominant heating mechanism, while C\textsc{ii} starts to dominate the cooling rate above $N_{\rm{H_{tot}}} \sim 10^{21} \, \text{cm}^{-2}$. It is also in this region that hydrogen becomes fully molecular, driven initially by an increase in the H\textsc{i} abundance and the rising density, and ultimately by self-shielding. Finally, dust shielding attenuates the radiation flux below 1 Ryd at column densities above $N_{\rm{H_{tot}}} \sim 10^{21} \, \text{cm}^{-2}$, which strongly cuts off the dust heating rate and also allows CO to form. In this fully shielded region heating is primarily from cosmic rays, while cooling is mostly from CO and H$_{2}$.

Finally, we compare the H\textsc{i}-H$_{2}$ transition predicted by our one-dimensional plane-parallel slab simulations in thermal and pressure equilibrium with two other prescriptions for molecular hydrogen formation that are already employed in hydrodynamic simulations: the simpler non-equilibrium model of \citet{gnedin09} \textit{(Gnedin09)}, and the analytic equilibrium model developed in \citet{krumholz08,krumholz09} and \citet{mckee10} \textit{(KMT)} (see figure~\ref{H2modelsFig}). 

At low pressure ($P / k_{\rm{B}} = 10^{3} \, \rm{cm}^{-3} \, \rm{K}$) the equilibrium H$_{2}$ fractions predicted by all three models generally agree well, as does the time evolution of the H$_{2}$ fraction predicted by our model and the Gnedin09 model. However, at low metallicity (0.1 $Z_{\odot}$) cosmic ray dissociation of H$_{2}$ reduces the H$_{2}$ fraction in the fully shielded region by a factor of two in our model, but cosmic ray dissociation is not included in the KMT or the Gnedin09 models. At high pressure ($P / k_{\rm{B}} = 10^{5} \, \rm{cm}^{-3} \, \rm{K}$) the H\textsc{i}-H$_{2}$ transition predicted by our model in chemical equilibrium occurs at a slightly lower column density than in the KMT model (e.g. $f_{\rm{H_{2}}} = 0.5$ at $N_{\rm{H_{tot}}} = 2.2 \times 10^{19} \, \rm{cm}^{-2}$ at solar metallicity, compared to $N_{\rm{H_{tot}}} = 3.0 \times 10^{19} \, \rm{cm}^{-2}$ in the KMT model). Furthermore, the H\textsc{i}-H$_{2}$ transition at this high pressure is flatter in the Gnedin09 model than in our model or the KMT model, due to the different H$_{2}$ self-shielding function that they use. 

\section*{Acknowledgments}
We thank Ewine van Dishoeck for useful discussions. We gratefully acknowledge support from Marie Curie Training Network CosmoComp (PITN-GA-2009- 238356) and from the European Research Council under the European Union's Seventh Framework Programme (FP7/2007-2013) / ERC Grant agreement 278594-GasAroundGalaxies.

{}

\appendix

\section{Shielding approximations}\label{shielding_approximations}

\begin{figure}
\centering
\mbox{
	\includegraphics[width=84mm]{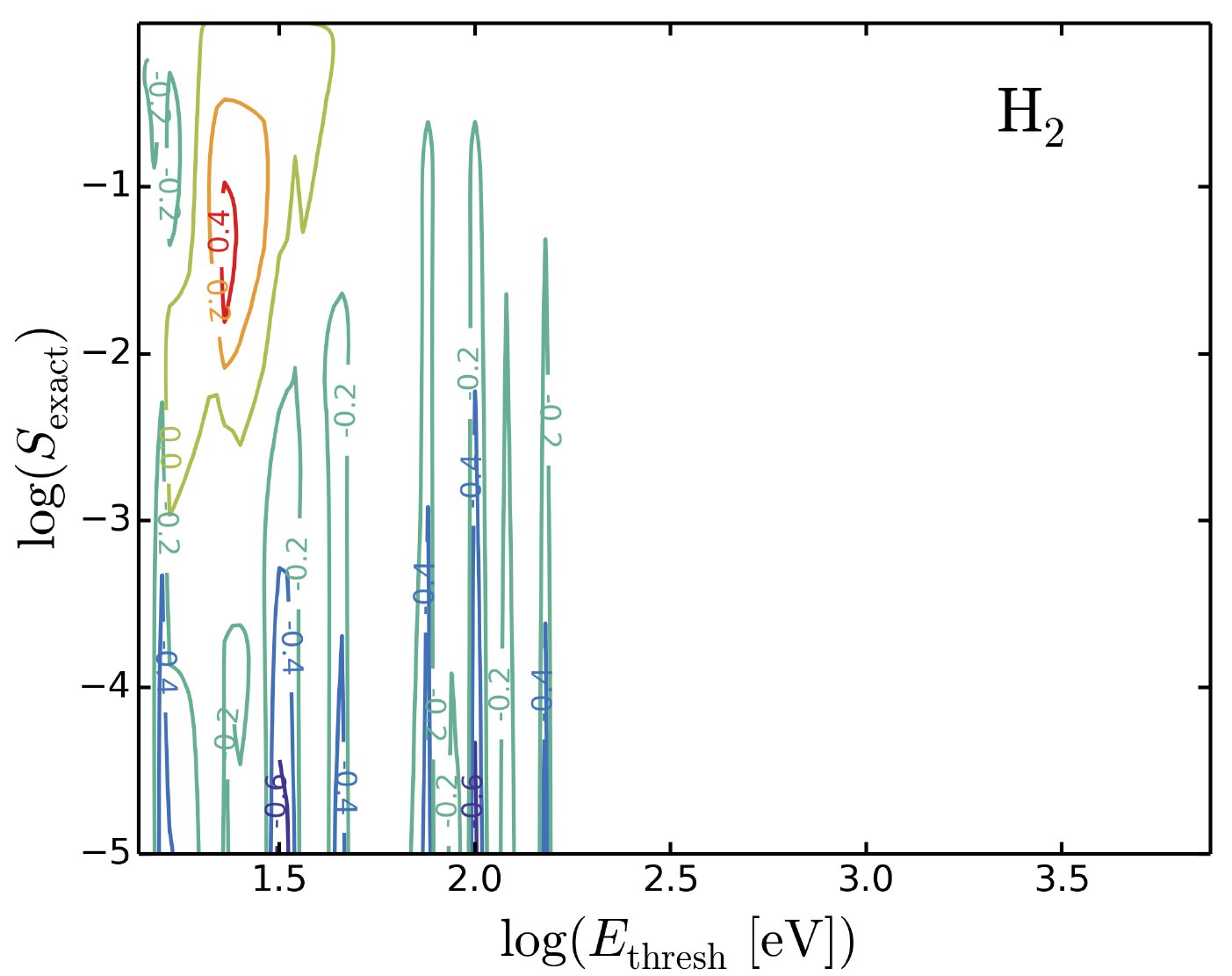}}
\mbox{
	\includegraphics[width=84mm]{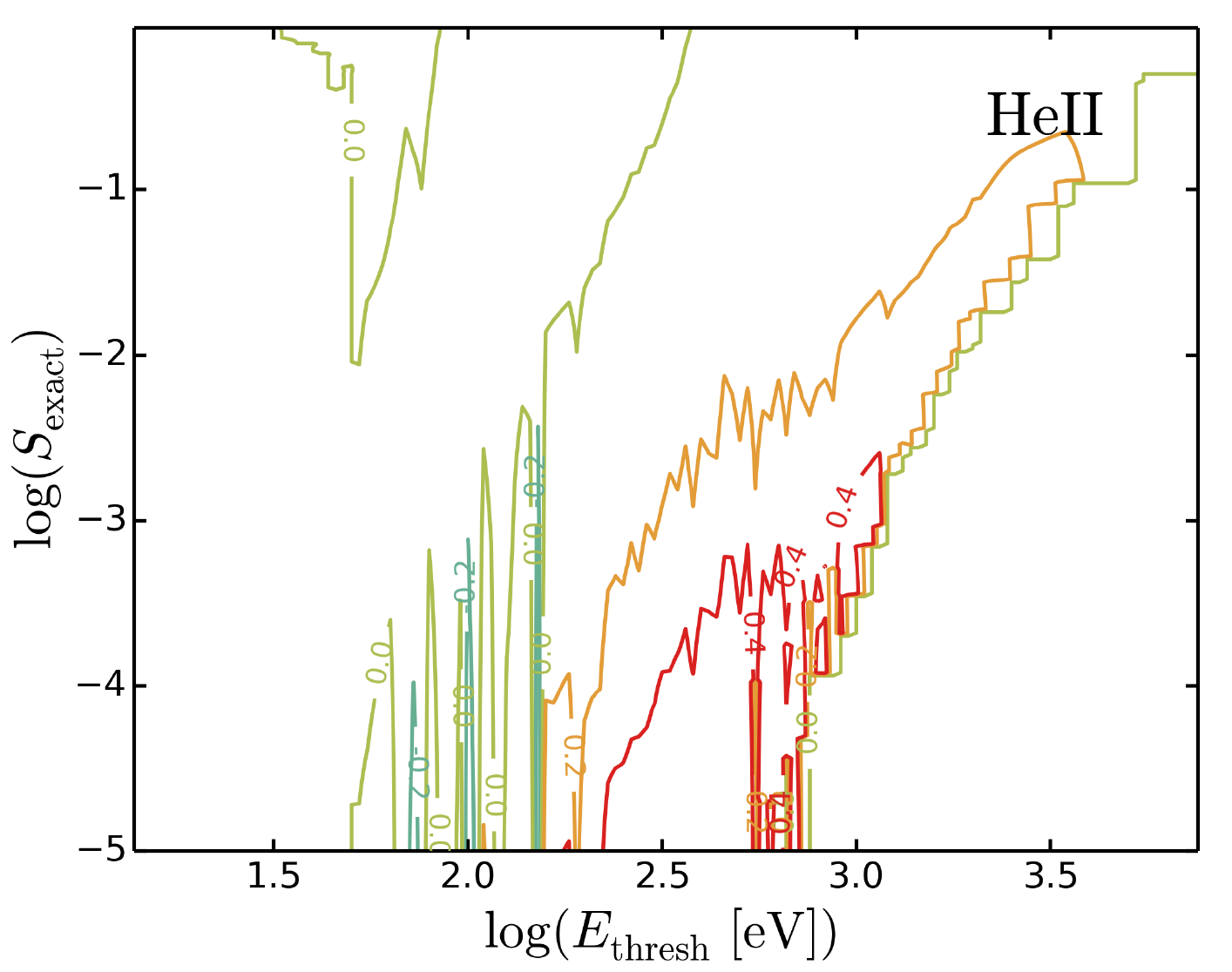}}
\caption{Contours of the relative errors in the ratio of optically thick to optically thin photoionisation rates of all species in our chemical network introduced by our approximations for absorption by H$_{2}$ and He\textsc{ii}, as given by equations~\ref{H2_approx_sigma}, \ref{HeII_approx_sigma} and \ref{photoIonThick_approx}, in the presence of the \citet{black87} interstellar radiation field. These contours are plotted against the threshold energy of each species, $E_{\rm{thresh}}$, on the x-axis, and the exact ratio of the optically thick to optically thin rates, $S_{\rm{exact}}$, on the y-axis. A value of $\log S_{\rm{exact}} = 0$ corresponds to the optically thin limit, and a value of $\log S_{\rm{exact}} = - \infty$ corresponds to the fully shielded limit. The top panel shows the errors if the radiation is absorbed purely by H$_{2}$ and the bottom panel shows if the radiation is absorbed purely by He\textsc{ii}. The largest errors that we see are $\sim 60 \%$, but these are rare and are typically found at $S_{\rm{exact}} \la 10^{-3}$, where the photoionisation rate is already strongly suppressed and so the errors are less important.}
\label{photoIonThickFig}
\end{figure}  

In section~\ref{shielding_section} we described how we calculate the photoionisation and photoheating rates after the radiation field has been attenuated by a total column density $N_{\rm{H_{tot}}}$. To compute the attenuation of the radiation by gas, we include absorption by H\textsc{i}, H$_{2}$, He\textsc{i} and He\textsc{ii}. However, to reduce the number of dimensions in which we tabulate the attenuated rates, H$_{2}$ and He\textsc{ii} absorption are included as follows. We assume that the H$_{2}$ cross section is three times the H\textsc{i} cross section above the H$_{2}$ ionisation energy (15.4 eV), and similarly that the He\textsc{ii} cross section is 0.75 times the He\textsc{i} cross section above the He\textsc{ii} ionisation energy (54.4 eV). In this section we show the errors that these approximations introduce to the photoionisation rates. 

For every species included in our chemical network, we calculate the ratio of the optically thick to optically thin photoionisation rates, $S_{\rm{gas}}$, from the \citet{black87} radiation field after it has been attenuated purely by H$_{2}$. We calculate this exactly using equation~\ref{photoIonThick}, then we calculate it with our approximation using equation~\ref{photoIonThick_approx}. In the top panel of figure~\ref{photoIonThickFig} we show contours of the relative error in $S_{\rm{gas}}$ that is introduced by our approximation for H$_{2}$ absorption, plotted against the threshold energy of each species, $E_{\rm{thresh}}$, on the x-axis, and the exact ratio of the optically thick to optically thin rates, $S_{\rm{exact}}$, on the y-axis. Blue contours indicate where our approximation underestimates the photoionisation rates and red contours indicate where we overestimate them. 

The largest errors seen in the top panel of figure~\ref{photoIonThickFig} are seen in the blue contours, which show that we underestimate the photoionisation rates of some species by up to $60 \%$ at $S_{\rm{exact}} \la 10^{-3}$. However, since these errors are found at low values of $S_{\rm{exact}}$, the photoionisation rates have already been suppressed and so are unlikely to be important in the chemical network. The more significant errors in this example are shown by the red contours in the top left corner of the top panel of figure~\ref{photoIonThickFig}. These show that we overestimate the photoionisation rates of species with $E_{\rm{thresh}} \sim 20$ eV by up to $40 \%$ at $S_{\rm{exact}} \sim 10^{-1}$. 

We also calculated the relative errors in the photoionisation rates when the radiation field is attenuated purely by He\textsc{ii}. These are shown in the bottom panel of figure~\ref{photoIonThickFig}. For species with $E_{\rm{thresh}} < 100$ eV, the errors are below $20 \%$. Larger errors are seen for species with higher threshold energies and mostly in the strongly shielded regime ($S_{\rm{exact}} \la 10^{-3}$). However, species with such high ionisation energies are unlikely to be found in shielded gas. 

The errors in the optically thick photoionisation rates shown in figure~\ref{photoIonThickFig} were calculated for absorption purely by H$_{2}$ or He\textsc{ii}. However, in practice the radiation will typically be absorbed by a combination of the four species that we consider in section~\ref{shielding_section}, including H\textsc{i} and He\textsc{i}, which we treat exactly. Therefore, figure~\ref{photoIonThickFig} represents upper limits on the actual errors. 

\section{H$_{2}$ self-shielding function}\label{H2self_comparison_section}

In this section we look at the accuracy of the H$_{2}$ self-shielding function that we use, as described in section~\ref{shieldDissoc_section}, by comparing it to the ratio of the optically thick to optically thin H$_{2}$ photodissociation rates predicted by \textsc{Cloudy} as a function of H$_{2}$ column density. We consider a plane-parallel slab of gas with primordial abundances that is illuminated from one side by the \citet{black87} interstellar radiation field. 

\begin{figure*}
\centering
\mbox{
	\includegraphics[width=160mm]{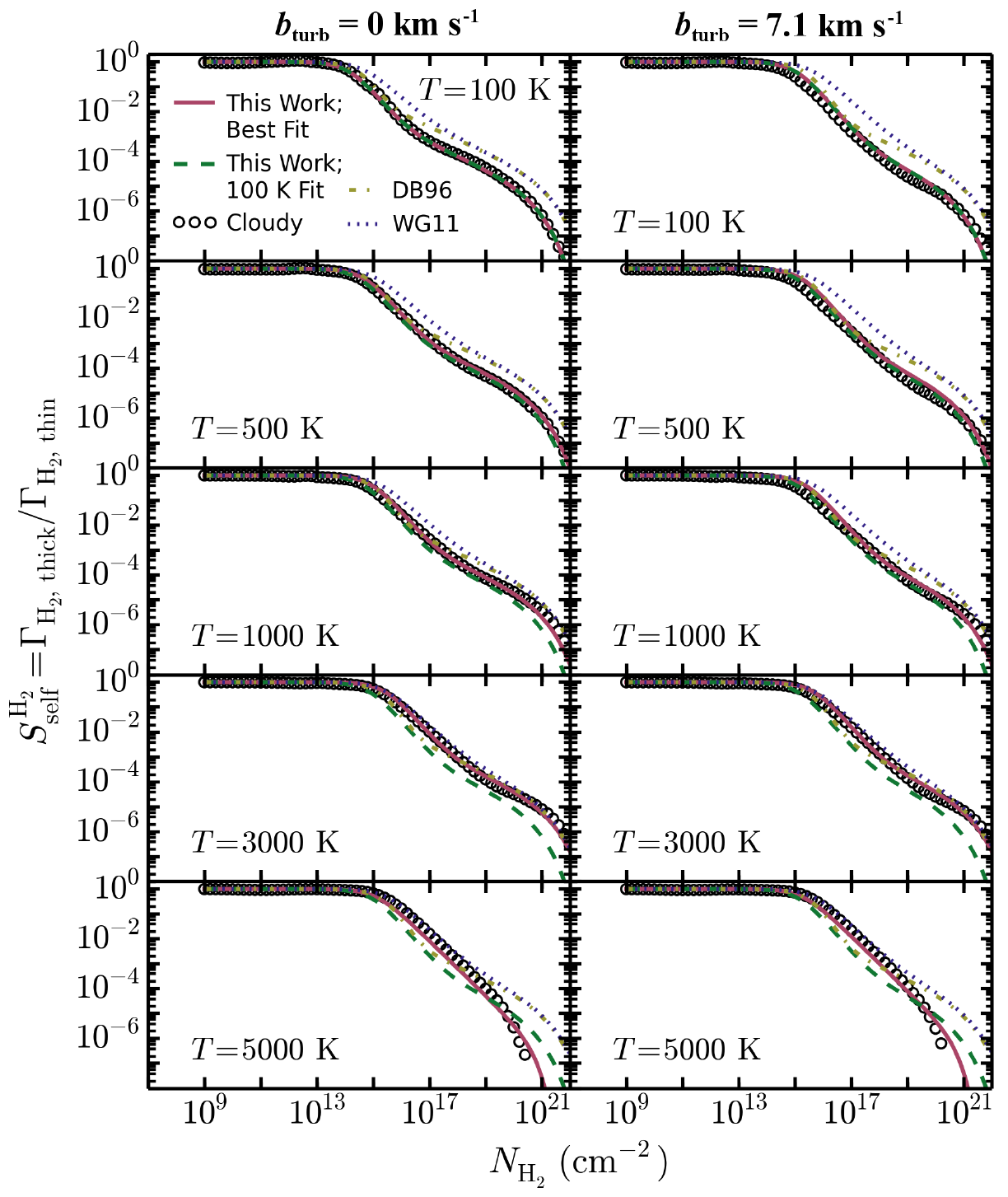}}
\caption{The ratio of the optically thick to optically thin H$_{2}$ photodissociation rates, $S_{\rm{self}}^{\rm{H_{2}}}$, plotted against the H$_{2}$ column density, $N_{\rm_{H_{2}}}$, as predicted by \textsc{Cloudy} \textit{(black circles)}, the self-shielding function of \citet{draine96} \textit{(DB96; yellow dot-dashed curves)}, the self-shielding function of \citet{wolcottgreen11} \textit{(WG11; blue dotted curves)}, our self-shielding function fitted to \textsc{Cloudy} at all temperatures \textit{(as given in equations~\ref{H2self_mod_eqn} to \ref{NcritEqn}; red solid curves)}, and our self-shielding function fitted to \textsc{Cloudy} at only 100 K \textit{(green dashed curves)}. Each row shows gas at different temperatures in the range $100 \, \rm{K} \leq T \leq 5000 \, \rm{K}$. In the left column Doppler broadening is purely thermal, while the right column includes turbulence with a Doppler broadening parameter of $b_{\rm{turb}} = 7.1 \, \rm{km} \, \rm{s}^{-1}$. In the absence of turbulence, the DB96 and WG11 functions overestimate $S_{\rm{self}}^{\rm{H_{2}}}$ by up to a factor $\sim 3$ at low temperatures, and they do not reproduce the temperature dependence of $S_{\rm{self}}^{\rm{H_{2}}}$ that is seen in \textsc{Cloudy}. Our best-fitting self-shielding function agrees with \textsc{Cloudy} to within 30 per cent at 100 K for $N_{\rm{H_{2}}} < 10^{21} \, \rm{cm}^{-2}$, and at 5000 K it agrees to within 60 per cent for $N_{\rm{H_{2}}} < 10^{20} \, \rm{cm}^{-2}$. The agreement between our best-fitting self-shielding function and \textsc{Cloudy} is poorer when turbulence is included, as it was fitted to the purely thermal Doppler broadening case. However, the DB96 and WG11 functions overestimate $S_{\rm{self}}^{\rm{H_{2}}}$ in cold gas by a larger factor when turbulence is included.}
\label{H2self_fig}
\end{figure*}

Figure~\ref{H2self_fig} shows the suppression factor of the H$_{2}$ photodissociation rate due to self-shielding, $S_{\rm{self}}^{\rm{H_{2}}}$, plotted against H$_{2}$ column density for five different temperatures in the range $100 \, \rm{K} \leq T \leq 5000 \, \rm{K}$ (shown in the different rows). The left panels were calculated using purely thermal Doppler broadening, while the right panels include turbulence with a Doppler broadening parameter $b_{\rm{turb}} = 7.1 \, \rm{km} \, \rm{s}^{-1}$. The black circles show the results from \textsc{Cloudy}, while the coloured curves show different analytic approximations to the H$_{2}$ self-shielding function. The yellow dot-dashed curves show the self-shielding function from \citet{draine96} (DB96), as given in equation~\ref{H2self_eqn} with $\omega_{\rm{H_{2}}} = 0.035$ and $\alpha = 2$, and the blue dotted curves show the suggested modification to this function given by \citet{wolcottgreen11} (WG11), with $\alpha = 1.1$. We see that, in the absence of turbulence, the WG11 function gives weaker self-shielding than the DB96 function at intermediate column densities ($10^{15} \, \rm{cm}^{-2} \la N_{\rm{H_{2}}} \la 10^{17} \, \rm{cm}^{-2}$), by up to a factor $\sim 6$. However, both of these functions underestimate the strength of the self-shielding in cold gas compared to \textsc{Cloudy}. For example, at 100 K the value of $S_{\rm{self}}^{\rm{H_{2}}}$ predicted by \textsc{Cloudy} is a factor of $\sim 3$ lower than the DB96 and WG11 functions at column densities $N_{\rm{H_{2}}} \ga 10^{17} \, \rm{cm}^{-2}$. 

To improve the agreement with \textsc{Cloudy}, we modified the self-shielding function of \citet{draine96} as shown in equation~\ref{H2self_mod_eqn}, then we fitted the parameters $\omega_{\rm{H_{2}}}(T)$, $\alpha(T)$ and $N_{\rm{crit}}(T)$ to the suppression factor $S_{\rm{self}}^{\rm{H_{2}}}$ predicted by \textsc{Cloudy} as a function of H$_{2}$ column density for each temperature separately, including only thermal broadening. To reproduce the temperature dependence of $S_{\rm{self}}^{\rm{H_{2}}}$ seen in \textsc{Cloudy} we then fitted analytic formulae to these three parameters as a function of temperature. These are given in equations~\ref{omegaEqn} to \ref{NcritEqn}. 

Our best-fitting self-shielding function is shown by the red solid curves in figure~\ref{H2self_fig}. In the absence of turbulence, this function agrees with \textsc{Cloudy} to within 30 per cent at $T = 100$ K for column densities $N_{\rm{H_{2}}} < 10^{21} \, \rm{cm}^{-2}$, and to within 60 per cent at $T = 5000$ K for $N_{\rm{H_{2}}} < 10^{20} \, \rm{cm}^{-2}$. 

To highlight the temperature dependence of $S_{\rm{self}}^{\rm{H_{2}}}$, we also show our best-fitting function using the parameters $\omega_{\rm{H_{2}}}(T)$, $\alpha(T)$ and $N_{\rm{crit}}(T)$ fixed at $T = 100$ K. These are indicated by the green dashed curves in figure~\ref{H2self_fig}. This represents the self-shielding function that we would get if we only fitted equation~\ref{H2self_mod_eqn} at 100 K. Note that the green curves in each panel of figure~\ref{H2self_fig} are not identical as they use the thermal Doppler broadening parameter for the given temperature. 

Comparing the red and the green curves in figure~\ref{H2self_fig}, we see that the self-shielding is weaker at temperatures $500 \, \rm{K} \la T \la 3000 \, \rm{K}$ when we use our full temperature-dependent function (red curves), compared to the function fitted only at 100 K (green curves). This agrees with the trend seen by \citet{wolcottgreen11}. However, at $T = 5000$ K the shape of the self-shielding function changes, and it becomes steeper than at lower temperatures for high column densities. This results in stronger self-shielding at $N_{\rm{H_{2}}} \ga 10^{19} \, \rm{cm}^{-2}$ compared to the fit at 100 K. 

Our self-shielding function was fitted to the predictions from \textsc{Cloudy} for gas with purely thermal Doppler broadening. To confirm that our best-fitting self-shielding function is also valid for turbulent gas, we repeated this comparison for gas that includes a turbulent Doppler broadening parameter $b_{\rm{turb}} = 7.1 \, \rm{km} \, \rm{s}^{-1}$. This comparison is shown in the right panels of figure~\ref{H2self_fig}. 

When we include turbulence, the agreement between our best-fitting self-shielding function and \textsc{Cloudy} is poorer. For example, at 100 K we overestimate $S_{\rm{self}}^{\rm{H_{2}}}$ by up to 90 per cent compared to \textsc{Cloudy} for $10^{15} \, \rm{cm}^{-2} < N_{\rm{H_{2}}} < 10^{20} \, \rm{cm}^{-2}$. However, this agreement is still much better than that between \textsc{Cloudy} and the DB96 and WG11 functions. For example, in the same range of column densities the WG11 function overestimates $S_{\rm{self}}^{\rm{H_{2}}}$ by up to an order of magnitude at 100 K.

\label{lastpage}

\end{document}